\documentclass[a4paper,11pt]{article}
\pdfoutput=1 

\usepackage{jcappub} 

\usepackage[varg]{txfonts}
\usepackage[utf8]{inputenc}

\usepackage{longtable}
\usepackage{epsfig,caption}
\usepackage{color}

\usepackage{lineno} 

\newcommand\ion[2]{#1$\,${\scshape{#2}}}
\newcommand{\lya}{Ly\ensuremath{\alpha}}%

\title{\bf  Constraints on neutrino masses from  Lyman-alpha forest
  power spectrum with BOSS and XQ-100 }

\author[a,b]{Christophe Y\`eche,}
\author[a]{Nathalie Palanque-Delabrouille,}
\author[a]{Julien Baur,}
\author[a]{H\'elion du Mas des Bourboux}

\emailAdd{christophe.yeche@cea.fr, nathalie.palanque-delabrouille@cea.fr, julien.baur@cea.fr,helion.du-mas-des-bourboux@cea.fr}

\affiliation[a]{CEA, Centre de Saclay, IRFU/SPP,  F-91191 Gif-sur-Yvette, France}
\affiliation[b]{Lawrence Berkeley National Laboratory, Berkeley, CA 94720, USA}

\date{Received xx; accepted xx}

\abstract{ 

We present constraints on  masses of active and sterile neutrinos in the context of the $\Lambda$CDM$\nu$ and $\Lambda$WDM models, respectively. We use   the  
one-dimensional Ly$\alpha$-forest power spectrum  from the Baryon
Oscillation Spectroscopic Survey (BOSS) of the Sloan Digital Sky
Survey (SDSS-III)  measured by Palanque-Delabrouille et al.~\cite{Palanque-Delabrouille2013}, and from the VLT/XSHOOTER legacy survey (XQ-100). In this paper, we present  our own measurement of the publicly released XQ-100 quasar spectra, focusing in particular on an improved determination of the  spectrograph resolution that allows us to push to smaller scales than the public release and reach $k$-modes of 0.070~$\rm s\,km^{-1}$. We compare the obtained  1D Ly$\alpha$ flux power spectrum  to the one measured by Irsic et al.~\cite{Irsic2016} to $k$-modes of $0.057\,\rm s\,km^{-1}$. 

Fitting Ly$\alpha$ data alone leads to cosmological parameters in excellent agreement with the values derived independently from Planck 2015 Cosmic Microwave Background (CMB) data. Combining BOSS  and XQ-100 Ly$\alpha$ power spectra, we  constrain the sum of neutrino masses to $\sum m_\nu < 0.8$~eV (95\% C.L.) including all identified sources of systematic uncertainties. With the addition of CMB data, this bound  is tightened to $\sum m_\nu < 0.14$~eV (95\% C.L.).

With their sensitivity to small scales, Ly$\alpha$ data are ideal to constrain $\Lambda$WDM models. Using 
 XQ-100 alone, we issue lower bounds on pure dark matter particles: $m_X \gtrsim 2.08 \: \rm{keV}$ (95\% C.L.) for early decoupled thermal relics, and $m_s \gtrsim 10.2 \: \rm{keV}$ (95\% C.L.)  for non-resonantly produced right-handed neutrinos.  Combining the 1D  Ly$\alpha$-forest power spectrum measured by BOSS and XQ-100, we improve the two bounds to $m_X \gtrsim 4.17 \: \rm{keV}$ and $m_s \gtrsim 25.0 \: \rm{keV}$ (95\% C.L.), slightly more constraining  than what was  achieved in Baur et al. 2015~\cite{Baur2016}  with BOSS data alone. The  $3~\sigma$ bound shows a more significant improvement, increasing from $m_X \gtrsim 2.74 \: \rm{keV}$ for BOSS alone to $m_X \gtrsim 3.10 \: \rm{keV}$ for the combined BOSS+XQ-100 data set. 
 	
Finally, we include in our analysis the first two redshift bins  ($z=4.2$ and $z=4.6$)  of the power spectrum measured by Viel et al. 2013~\cite{Viel2013} with the high-resolution HIRES/MIKE spectrographs. The addition of HIRES/MIKE power spectrum allows us to further improve the two limits to  $m_X \gtrsim 4.65 \: \rm{keV}$ and $m_s \gtrsim 28.8  \: \rm{keV}$ (95\% C.L.).

}

\begin{document}
\maketitle
\flushbottom

\section{Introduction}
\label{sec:intro}
The flux power spectrum of the Lyman-$\alpha$ (Ly$\alpha$) forest in quasar absorption spectra is a powerful tool to study clustering in the  Universe at redshifts  $\sim$2 to 4, on scales ranging from a few Mpc~\cite{Croft2002,McDonald2006,Palanque-Delabrouille2013} to hundreds of Mpc~\cite{Slosar2013,Busca2012,Delubac2015}. Compared to a model derived from a set of dedicated hydrodynamical simulations, the Ly$\alpha$-flux power spectrum can provide valuable information on the formation of structures and their evolution. In particular, by probing scales down to a few Mpc, the 1D flux power spectrum is  sensitive to neutrino masses through the suppression of power on small scales that neutrinos induce. Being  relativistic until late in the history of the Universe, neutrinos free-stream out of gravitational potentials and therefore damp  small-scale  density fluctuations for at least two reasons: by the absence of neutrino perturbations in the total matter power spectrum, but even more so by slowing down the  growth rate  of baryons and CDM perturbations since neutrinos contribute to the background density, and thus to the  expansion rate, but not to the clustering. The overall effect of massive neutrinos is a step-like suppression of power by a factor of order $ 1 - 8(\Omega_\nu/\Omega_m)$, where the present neutrino energy density relative to the critical density is given  by $\Omega_\nu = \sum m_\nu/(93.14\, h^2\, \rm eV^2)$~\cite{Seljak2006,Lesgourgues:2012uu}.

Cosmic Microwave Background (CMB) data can also constrain $\sum m_\nu$. In the standard thermal history of the Universe, massless neutrinos have a temperature  $T_\nu =  0.18$~eV at the epoch of last scattering, corresponding to an average momentum $\langle p \rangle = 3.15 \, T_\nu = 0.57$~eV. This temperature sets the range of masses for which neutrinos start to have an appreciable effect on the CMB power spectrum to $\sum m_\nu > 3\times 0.57 = 1.7~{\rm eV}$ (for three active neutrino species). Below this mass, the neutrinos are still relativistic at recombination and have no direct impact on the primary CMB anisotropies. The effect of neutrino mass on the  CMB then only appears at the level of secondary anisotropies, through the integrated Sachs-Wolf effect or the weak lensing by foreground gravitational structures. Using a measurement of these effects, the latest limit set  on $\sum m_\nu$ by the Planck team from CMB data alone is at the level of 0.7~eV~\cite{Planck2015}. 

Despite a clear suppression in the power spectrum, Ly$\alpha$ data alone also exhibit a sensitivity to $\sum m_\nu$ at the level of about 1~eV only, due to the fact that the scales probed by Ly$\alpha$ forests are  in a region of scales where the ratio of the power spectra for massive to massless neutrinos is quite flat. However, a tight constraint on $\sum m_\nu$ can be obtained by combining  Ly$\alpha$ with CMB data: while Ly$\alpha$ data  probe the suppressed power spectrum, CMB data on the other hand probe the power spectrum on scales large enough to be unaffected by the free-streaming of neutrinos. The combination of data from small and large scales therefore provides a direct measure of the suppression, and thus on $\sum m_\nu$.  In practice, Ly$\alpha$ measures the power spectrum level, defined by $\sigma_8$ and  $\Omega_m$,  CMB provides the correlations between these parameters and $\sum m_\nu$, and the joint use of these two probes significantly improves the constraint on $\sum m_\nu$ compared to what  either probe alone can achieve.

By interfering with the gravitational collapse of structures while they are relativistic, warm dark matter particles also significantly affect the matter power spectrum and can  be studied in a similar way as active neutrinos.  
In the case of $\Lambda$WDM models where all the dark matter is assumed to be in the form of WDM particles with masses of a few keV, the linear 3D matter power spectrum shows a complete cut-off on  scales above $k \sim 2\pi/r \sim 10\, h\,{\rm Mpc^{-1}}$~\cite{VLH05,SMT08,VBH08}. This free-streaming-induced cut-off  translates into a gradual suppression on the 1D flux power spectrum that falls within  the range of scales probed by the Ly$\alpha$ forest of distant high redshift quasars. Ly$\alpha$ forest data therefore again provide  an ideal tool to study keV-range WDM and yield lower bounds on the mass of early decoupled thermal relics or right-handed sterile neutrinos, for instance. 

In this paper, we exploit  Ly$\alpha$ data  from three surveys. For the Baryon Oscillation Spectroscopic Survey (BOSS) of the Sloan Digital Sky Survey (SDSS-III), we  use the 1D Ly$\alpha$  flux power spectrum measured by~\cite{Palanque-Delabrouille2013} with the quasar data of the DR9 release. For VLT/XSHOOTER legacy survey XQ-100~\cite{Lopez2016}, we directly compute our own 1D power spectrum from the publicly released XQ-100 quasar spectra. A comparison with the power spectrum measured by~\cite{Irsic2016} is  discussed in this paper.  Finally we add to the BOSS+XQ-100 data set, the power spectrum measured for two redshift bins ($z=4.2$ and $z=4.6$) with the high-resolution HIRES/MIKE spectrographs and described in~\cite{Viel2013}.

The simulations we use to interpret the Ly$\alpha$ data come from a grid of 36 hydrodynamical simulations having a resolution equivalent to $3\times 3072^3$ particles in a $(100~h^{-1}~{\rm Mpc})^3$ box \cite{Borde2014,Rossi2014}. We use these simulations to predict the flux power spectrum and  constrain  cosmology, the sum of the neutrino masses $\sum m_\nu$ and the mass of WDM particles. 

The layout of the paper is as follows. Sec.~\ref{sec:xq100} presents the measurement of the Ly$\alpha$ forest power spectrum  from the VLT/XSHOOTER legacy survey XQ-100 data. We explain the various steps of the analysis and detail the  non-standard issues that we addressed with specific care, such as the determination of the spectrograph resolution in the present case where the seeing of the observations is smaller than the slit size.   Sec.~\ref{sec:data} gives a brief summary of the Ly$\alpha$ forest, CMB and Baryon Acoustic Oscillation (BAO)  data sets that we use in this work.  We also introduce the hydrodynamical simulations from which we build our likelihood. The main objective of Sec.~\ref{sec:LyaAlone} is to present the cosmological constraints that can be achieved from  Ly$\alpha$ data alone (BOSS and XQ-100). The base model we consider is a flat $\Lambda$CDM cosmology with massive neutrinos, thereafter referred to as the base $\Lambda$CDM$\nu$ cosmology.  In Sec.~\ref{sec:LyaCMB}, we include additional data, namely several configurations of CMB  and BAO measurements. We present the results obtained on the parameters of our base $\Lambda$CDM$\nu$ cosmology using various combinations of the data sets. Finally, in Sec.~\ref{sec:WDM}, we discuss our results in the framework of $\Lambda$WDM models and give lower limits on the mass of thermal relics and  non-resonantly produced sterile neutrino.  

This paper refers extensively to the earlier papers that reported constraints on cosmological parameters and  the  mass of active neutrinos~\cite{Palanque-Delabrouille2015} or constraints on WDM and the mass of sterile neutrinos~\cite{Baur2016} using Ly$\alpha$ data from the SDSS-III/BOSS survey. For the sake of simplicity, we will henceforth refer to~\cite{Palanque-Delabrouille2015} as PY15 and to \cite{Baur2016} as BP16. We also refer the reader to \cite{Borde2014} for a detailed description of the grid of hydrodynamical simulations used in this work, and to \cite{Rossi2014} for the implementation of neutrinos  and their impact on the 1D flux power spectrum. Definitions of the most relevant symbols used in this paper can be found in Tables~\ref{tab:astroparam} and~\ref{tab:cosmoparam}.

\begin{table}[htbp]
\caption{Definition of astrophysical parameters}
\begin{center}
\begin{tabular}{p{2.9cm}ll}
\hline
\hline
Parameter& Definition \\
\hline
$\delta=\rho /  \left\langle \rho \right\rangle$  \dotfill  & Normalized baryonic density $\rho$ of IGM\\ 
$T$ \dotfill & Temperature of IGM modeled by $T=T_0  \cdot  \delta^{\gamma-1}$ \\
$T_0$ \dotfill & Normalization temperature of IGM at $z=3$ \\
$\gamma$ \dotfill &  Logarithmic slope of  $\delta$ dependence of IGM temperature  at $z=3$ \\
$\eta^{T_0}$ \dotfill &  Logarithmic slope of  redshift dependence of $T_0$ (different for $z<$ or $>3$) \\
$\eta^{\gamma}$  \dotfill &  Logarithmic slope of  redshift dependence of $\gamma$ \\
$A^\tau$  \dotfill &  Effective optical depth of Ly$\alpha$ absorption  at $z=3$\\
$\eta^\tau$  \dotfill &  Logarithmic slope of  redshift dependence of $A^\tau$   \\
$f_ {\rm{Si\,III}}$  \dotfill &  Fraction of Si\,III absorption relative to Ly$\alpha$ absorption\\
$f_ {\rm{Si\,II}}$  \dotfill &  Fraction of Si\,II absorption relative to Ly$\alpha$ absorption\\
\hline
\end{tabular}
\end{center}
\label{tab:astroparam}
\end{table}%

\begin{table}[htbp]
\caption{Definition of cosmological  parameters}
\begin{center}
\begin{tabular}{p{4cm}l}
\hline
\hline
Parameter&  Definition \\
\hline
$\Omega_m$ \dotfill & Matter fraction today (compared to critical density) \\
$H_0$ \dotfill &  Expansion rate today in km s$^{-1}$ Mpc$^{-1}$ \\
$\sigma_8$ \dotfill & RMS matter fluctuation amplitude today in linear theory \\
$\tau$ \dotfill & Optical depth to reionization\\
$z_{\rm re}$ \dotfill & Redshift where reionization fraction is 50\%\\
$n_s$ \dotfill & Scalar spectral index \\
$\sum m_\nu$ \dotfill &  Sum of neutrino masses in eV \\
$m_X$ \dotfill &   Mass of thermal relics in keV \\
$m_s$ \dotfill &   Mass of  non-resonantly produced sterile neutrino in keV \\
\hline
\end{tabular}
\end{center}
\label{tab:cosmoparam}
\end{table}%


\section{XQ-100 1D power spectrum }
\label{sec:xq100}
In this section we present our measurement of the one-dimensional Ly$\alpha$ forest power spectrum from the quasars of the XQ-100 project. After a brief description of the data, we explain the analysis of the spectra, building upon   the method developed for  BOSS and described in~\cite{Palanque-Delabrouille2013}. Finally, we compare our results to the measurement of the power spectrum from the same data set by Irsic et al.~\cite{Irsic2016}.

\subsection{XQ-100 survey}

XQ-100, ``Quasars and their absorption lines: a legacy survey of the high-redshift universe with VLT/XSHOOTER'',  is one of the large programmes of the European Southern Observatory~\cite{Lopez2016}.  The survey consists of  a homogeneous and high-quality sample of 100 echelle spectra of  quasars at redshifts $z \simeq 3.5-4.5$. The quasars were  observed with full spectral coverage from $3\,150$ to $25\,000\,\AA$, at a resolving power ranging from $ \sim  4\,000$ to $7\,000$, depending on  wavelength.

 We use  both the XQ-100 raw data and the XQ-100 Science Data Products (SDP) released at \url{http://archive.eso.org/eso/eso_archive_main.html} and \url{http://archive.eso.org/wdb/wdb/adp/phase3_main/form}, respectively. The main part of our work relies on the SDP spectra; the raw spectra are only exploited for the study and  validation of the spectrograph resolution, see Sec.~\ref{sec:reso}.
 
The region of interest for this study, the Ly$\alpha$ forest,  is covered by the UVB $(3\,150-5\,600\,\AA)$ and VIS $(5\,400-10\,200\,\AA)$ spectroscopic arms of XSHOOTER.  In the  Ly$\alpha$ forest, the signal-to-noise ratio per pixel varies from 5 to 60, with  an average of $\sim 25$. The spectral resolution depends on the arm. It is respectively $12\,{\rm km\,s^{-1}}$ and  $ 18\,{\rm km\,s^{-1}}$ on average for the VIS and UVB arms respectively.

For comparison,  about $700$ BOSS quasars  in the same HI absorption region ($z \simeq 3.0-4.2$) were analyzed in~\cite{Palanque-Delabrouille2013}.  BOSS quasars   exhibit  a signal-to-noise ratio typically 10 to 20 times lower and  a spectral resolution 3 to 5 times worse than XQ-100 quasars.  As a consequence,  BOSS data allow us to compute the power spectrum to  scales at most of $0.02\,{\rm s\,km^{-1}}$, while XQ-100 data allow us to reach much smaller scales, corresponding to $0.07\,{\rm s\,km^{-1}}$ for the VIS arm of XSHOOTER that has the better resolution.

\subsection{The normalized transmitted flux fraction  \texorpdfstring{$\delta(\lambda)$}{delta}  }

\begin{figure}[htbp]
\begin{center}
\epsfig{figure= 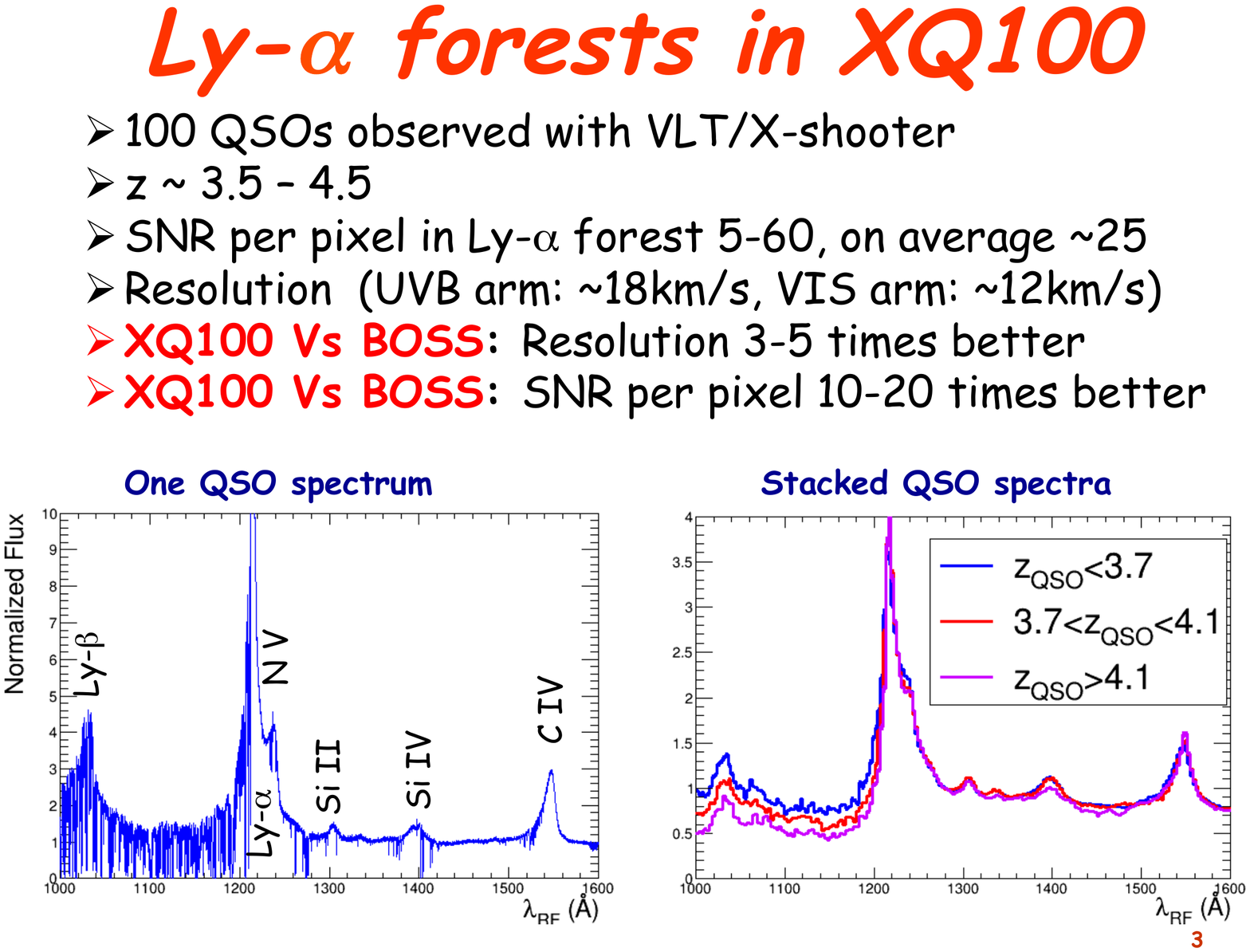,width = 7.2cm}
\epsfig{figure= 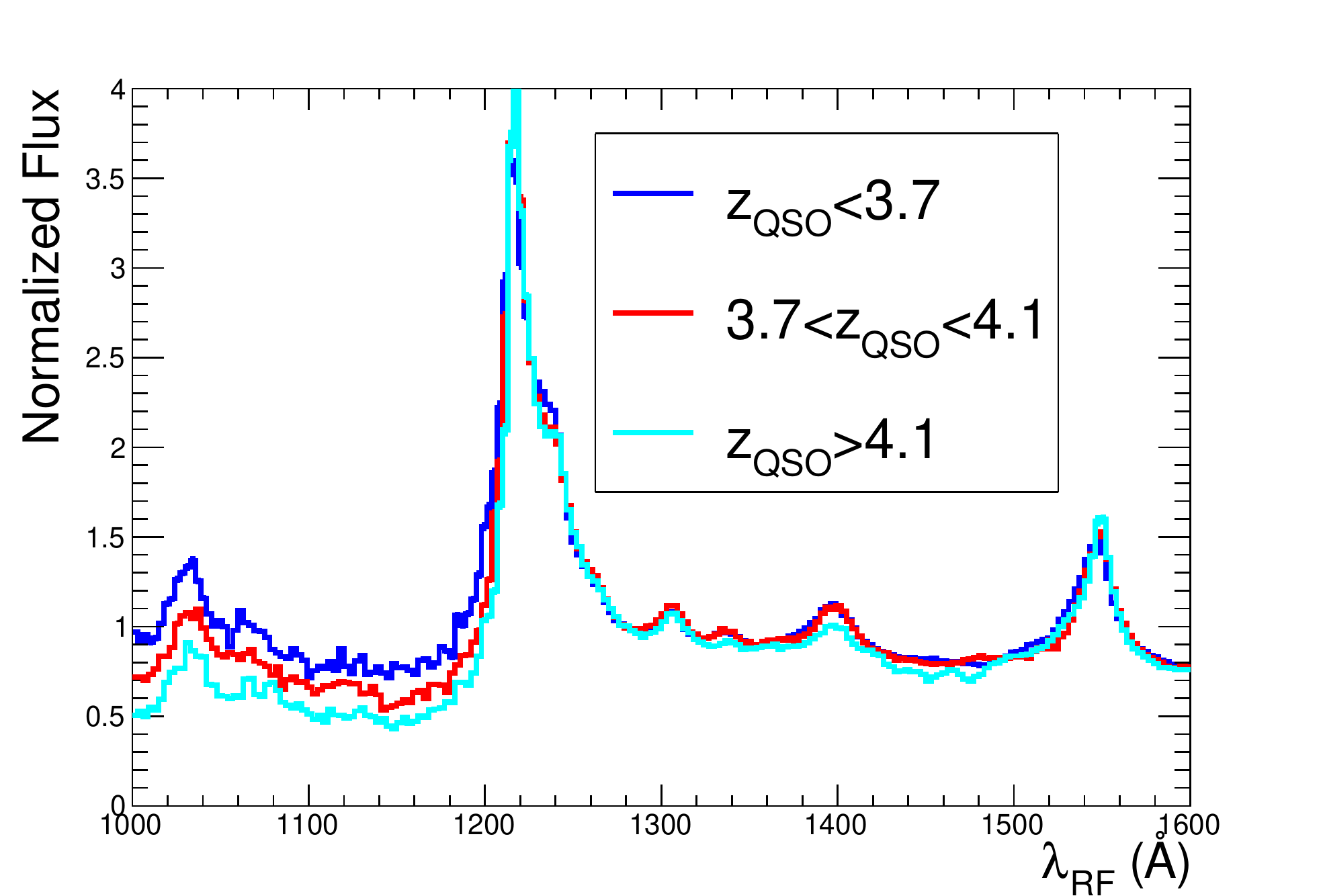,width = 7.8cm}
\caption{\it Left: Example of a typical quasar spectrum observed by XQ-100. Right:   Average quasar spectra in three redshift bins. All spectra are normalized at $\lambda = 1280$\,\AA. } 
\label{fig:SpectraXQ100}
\end{center}
\end{figure}

In Fig.~\ref{fig:SpectraXQ100}, we show  a typical XQ-100 spectrum (left plot) and  the  average quasar spectrum (right plot) obtained by averaging  all  XQ-100 quasar spectra  split into 3 redshift bins. Broad quasar emission lines are clearly visible, such as Ly$\beta$  ($1026\,$\AA), Ly$\alpha$  ($1216\,$\AA), \ion{N}{v}   ($1240\,$\AA), \ion{Si}{iv}   ($1400\,$\AA) and \ion{C}{iv}  ($1549\,$\AA), where all wavelengths are expressed in rest frame.  Absorption by Ly$\alpha$ absorbers along a quasar line of sight appears blueward of the quasar Ly$\alpha$ emission peak, with more absorption (and hence less transmitted flux) at high redshift. 

We define the Ly$\alpha$ forest by the range $1040<\lambda_{\rm RF}<1200\,$\AA, thus  $\sim4000\,{\rm km\,s^{-1}}$ away from the quasar Ly$\beta$ and Ly$\alpha$ emission peaks to avoid contamination of the power spectrum by astrophysical effects in the vicinity of the quasar. The Ly$\alpha$ forest spans a  redshift range $\Delta z \sim 0.65$ for a quasar at a redshift $z_{\rm qso}=4$.  In order to improve the redshift resolution,  we split the  Ly$\alpha$ forest into  three  consecutive and non-overlapping sub-regions of  equal  length, hereafter called `$z$-sectors'.  The splitting of each spectrum is done in such a way as to ensure that a  '$z$-sector' lies on a single spectroscopic arm  (either VIS or UVB), and to avoid the parts of the forest  containing   Damped Lyman Alpha (DLA)  or Lyman limit (LLS) systems detected  by~\cite{Sanchez-Ramirez2016}.

The  largest possible mode is determined by the Nyquist-Shannon limit at $k_{\rm Nyquist} = \pi/\Delta v$. With a pixel size $\Delta v = c\Delta \lambda/\lambda = 20\;{\rm km\,s^{-1}}$ and $ 11\;{\rm km\,s^{-1}}$ for the UVB and VIS arms, the largest mode is, respectively, $k_{\rm Nyquist} = 0.16\;{\rm s\,km^{-1}}$ and  $0.29\;{\rm s\,km^{-1}}$.  We limit the analysis, however,  to $k_{\rm max} = 0.05-0.07 \;{\rm s\,km^{-1}}$, depending on the absorption redshift, because of the large window function correction, at  the largest $k$-modes, mostly due to  the spectrograph resolution. 

Sky lines  affect the data quality by  increasing  the pixel noise. We identify major sky lines (such as lines at 5577, 5890, 6300, 6364, and $6864\,\AA$)  and  we replace the flux of each  pixel impacted by a sky line by the average value of the flux over the rest of the forest. This procedure introduces a small $k$-dependent bias in the resulting power spectrum, which is negligible in this analysis due to the small number of sky lines in this wavelength region and the good spectrograph resolution. 

The normalized transmitted flux  fraction $\delta(\lambda)$ is estimated from the pixel flux $f(\lambda)$ by:
\begin{equation}
\delta(\lambda) = \frac{f(\lambda)}{f_{\rm qso}^{1280}  C_q(\lambda,z_{\rm qso}) \overline{F}(z_{\rm Ly\alpha})} - 1\, ,
\label{eq:delta}
\end{equation}
where $f_{\rm qso}^{1280}$ is a normalization  equal to the mean flux in a $20\,$\AA\ window centered on $\lambda_{RF} = 1280\,$\AA\ where $\lambda_{RF}=\lambda/(1+z_{\rm qso})$,  $C_q(\lambda,z_{\rm qso})$ is the normalized unabsorbed flux (the mean quasar `continuum') and $\overline{F}(z_{\rm Ly\alpha})$ is the mean transmitted flux fraction at the \ion{H}{i} absorber redshift. Pixels affected by sky line emission are not included when computing the normalization. Since the mean quasar continuum is flat in the normalization region, the rejection of a few pixels does not bias the mean pixel value. The product $ C_q(\lambda,z_{\rm qso}) \overline{F}(z_{\rm Ly\alpha})$ is assumed to be universal for all quasars at redshift $z_{\rm qso}$ and is computed by stacking appropriately-normalized quasar spectra $f/f_{\rm qso}^{1280}$, thus averaging out the fluctuating Ly$\alpha$ absorption.  The product $f_{\rm qso}^{1280}  C_q(\lambda,z_{\rm qso}) \overline{F}(z_{\rm Ly\alpha})$ represents the mean expected flux, and   the transmitted flux fraction is given by $F=f / (f_{\rm qso}^{1280}C_q)$. For a pixel at wavelength $\lambda$, the corresponding \ion{H}{i} absorber redshift $z_{\rm Ly\alpha}$ can be inferred from $1+z_{\rm Ly\alpha} = \lambda / \lambda_{\rm Ly\alpha}$, where $\lambda_{\rm Ly\alpha} = 1215.67$\AA.

\subsection{Discussion of spectrograph resolution}
\label{sec:reso}

In the analysis of BOSS data~\cite{Palanque-Delabrouille2013}, we  encountered two main issues:  determination of  noise power spectrum $P^{noise}(k)$ and  correction of the spectrograph resolution. As the signal-to-noise ratio per pixel is much better in XQ-100 than in BOSS and since the data are at higher redshift, we can anticipate that with XQ-100  the impact of  $P^{noise}(k)$ will be negligible. On the other hand, we want to study the small scales that are key for  constraints on WDM. We aim at exploiting scales  to $0.07\,{\rm s\,km^{-1}}$, three times smaller than with BOSS. As the spectral resolution of XSHOOTER  is 3 to 5 times better than for BOSS, the knowledge and  control of the spectrograph resolution is again a key issue of the analysis. 

\begin{figure}[htbp]
\begin{center}
\epsfig{figure= 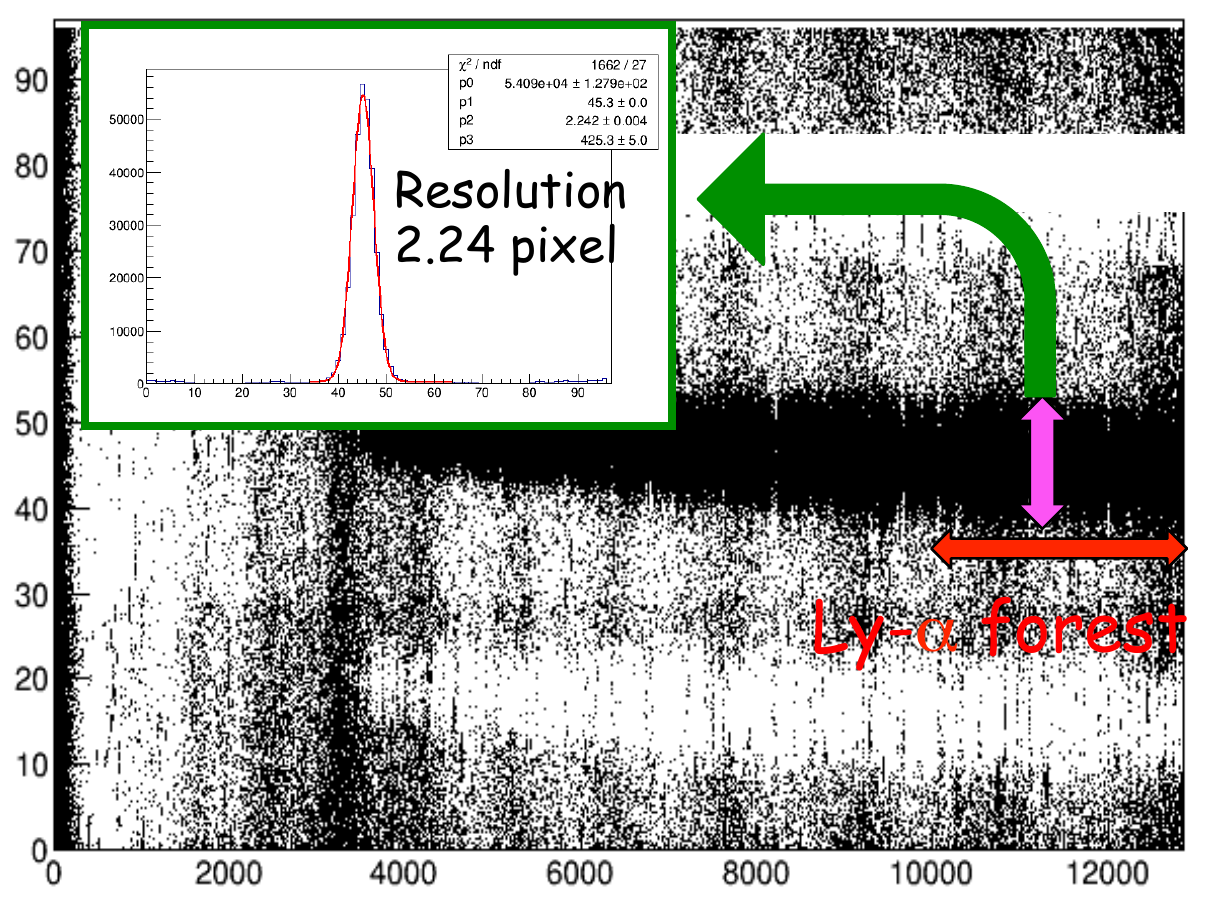,width = 7.0cm}
\epsfig{figure= 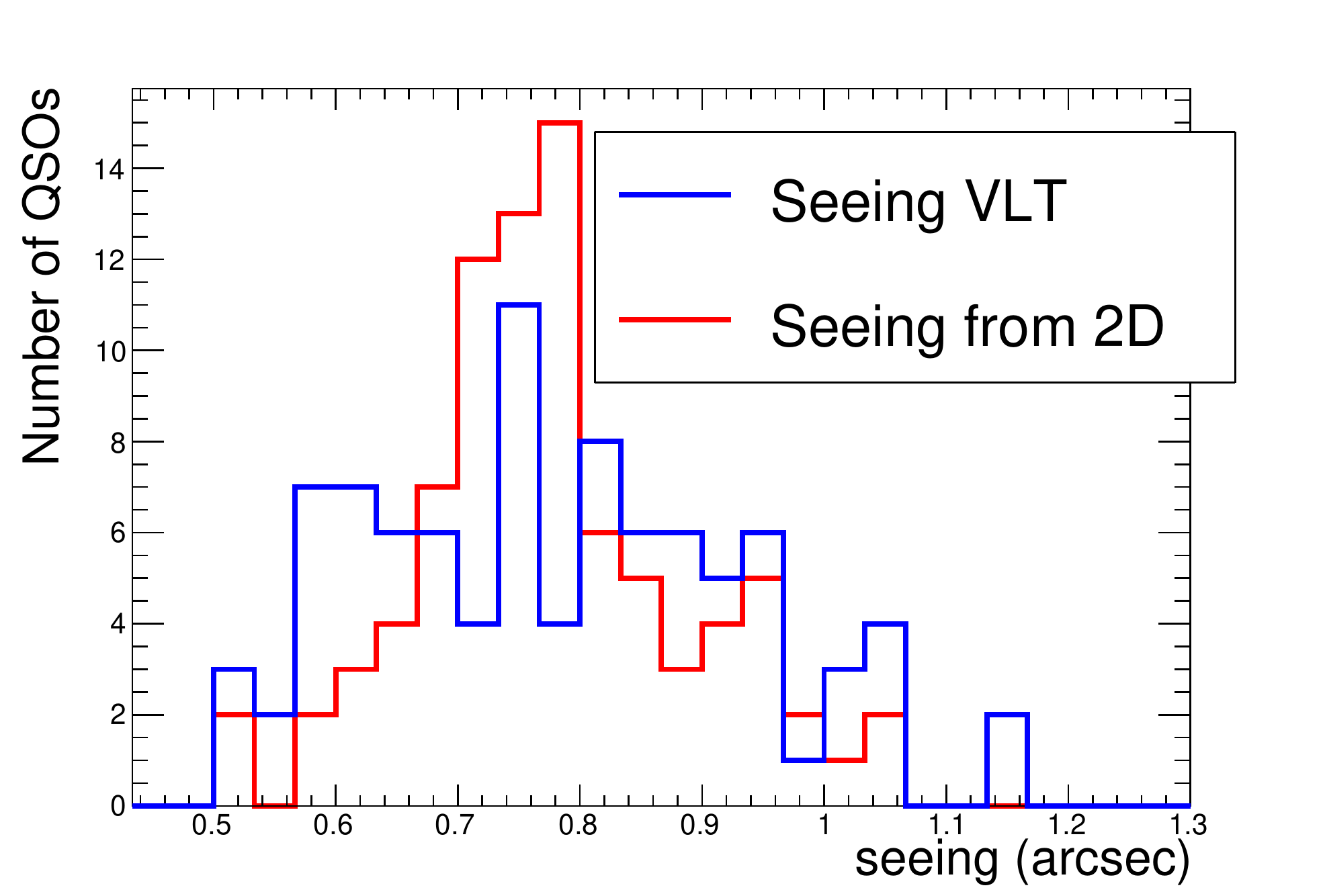,width = 8.0cm}
\caption{\it Left plot: example of a 2D spectrum for the UVB arm. The red arrow marks the position of the Ly-$\alpha$ forest. The transversal distribution along the purple arrow (i.e. orthogonal to the direction of the slit) allows us to visualize the PSF. The insert represents this distribution with a Gaussian fit to measure the seeing. Right plot: Distribution of the seeings for the quasars observed by XQ-100. The blue curve corresponds to the VLT measurements of the seeing during the spectroscopic observation. The red curve is derived from the 2D spectra. } 
\label{fig:XQ100Seeing}
\end{center}
\end{figure}

The spectral resolution of the UVB and VIS arms are derived from the slit widths, which were respectively $1.0''$ and $0.9''$. These slit widths provide a nominal resolving power of 4350 and 7450 as explained in~\cite{Lopez2016}. However, for many observations, the seeing was smaller than the slit width, inducing an underestimate of the resolving power. To address this issue, we first determine the seeing with the raw spectra of XQ-100. The 2D spectra, as shown on Fig.\ref{fig:XQ100Seeing} (left plot), allow us to measure the seeing by fitting a Gaussian of the transversal distribution (i.e. orthogonal to the direction of the slit). We checked that the seeing thus measured is in good agreement with the seeing measured simultaneously at VLT (see Fig.\ref{fig:XQ100Seeing}, right plot). On average the seeing is equal to $0.7''-0.8''$, better than the slit widths used during the observations.

As a consequence, we  decided to compute our own determination of the spectral resolution, using the VLT seeing, namely (SEEING\_MIN+SEEING\_MAX)/2 from   XQ-100\_summary.fits file. We  determine the  resolution from the tables describing the XSHOOTER instrument (defined at~\cite{XSHOOTER}). When the seeing is smaller than the slit width, we substitute the slit width by the computed seeing, and compute the corresponding  resolution.

\subsection{Computation of the  $P_{1D}(k)$}

\subsubsection{Method}

To measure the one-dimensional power spectrum $P_{1D}(k)$  we decompose each absorption spectrum $\delta_{\Delta v}$ into Fourier modes  and  estimate their variance as a function of wave number.  In practice, we do this  by computing the discrete Fourier transform of  the  flux transmission fraction $\delta=F/\langle F\rangle - 1$  as described in~\citet{Croft1998}, using a fast Fourier Transform  (FFT) algorithm. In~\cite{Palanque-Delabrouille2013}, we  developed in parallel a likelihood approach, in a very similar way to~\cite{McDonald2006}. We  demonstrated that the latter method is more appropriate when  noise and  resolution vary from one pixel to another and when many pixels are masked. Because of the excellent quality of the spectrograph of XSHOOTER, a simple FFT approach can be pursued for the present  analysis.  

The use of a FFT requires  the pixels to be equally spaced. The condition is satisfied with the  SDP spectra provided by the XQ-100 pipeline~\cite{Lopez2016}: the spectra are computed with a constant pixel  width  $\Delta[\log(\lambda)] $, and the velocity difference between pixels, i.e., the relative velocity of absorption systems at wavelengths $\lambda+\Delta\lambda/2$ and $\lambda-\Delta\lambda/2$, is $\Delta v =  c\, \Delta \lambda / \lambda = c\, \Delta [\ln(\lambda)]$. Throughout this paper we  therefore use velocity instead of observed wavelength. Similarly, the wave vector $k\equiv 2\pi/ \Delta v$ is measured in $\rm s\,km^{-1}$.

In the absence of instrumental effects (noise and resolution of the spectrograph), the one-dimensional power spectrum can be simply written as the ensemble average over quasar spectra of $P^{raw}(k) \equiv   \left| \mathcal{F}(\delta_{\Delta v})  \right|^2$, where $\mathcal{F}(\delta_{\Delta v})$ is the Fourier transform of the normalized flux transmission fraction $\delta_{\Delta v}$  in the quasar Ly$\alpha$ forest, binned in pixels of width $\Delta v$.

When taking into account the noise in the data, the impact of the spectral resolution of the spectrograph, the cross-correlated background  due to absorption by Ly$\alpha$ and \ion{Si}{iii} and the uncorrelated background due to metal absorption such as  \ion{Si}{iv}  or  \ion{C}{iv}, the raw power spectrum is
\begin{equation}
P^{raw}(k)= \left( P^{{\rm Ly}\alpha}(k) +   P^{{\rm Ly}\alpha-{\rm Si_{III}}}(k) +  P^{metals}(k)\right) \cdot W(k,R,\Delta v) + P^{noise}(k)
\label{eq:P1D_raw}
\end{equation}
where $W^2(k,R,\Delta v)$ is the window function related to spectrograph  resolution.

 \subsubsection{Correction of the instrumental effects}
 
\begin{figure}[htbp]
\begin{center}
\epsfig{figure= 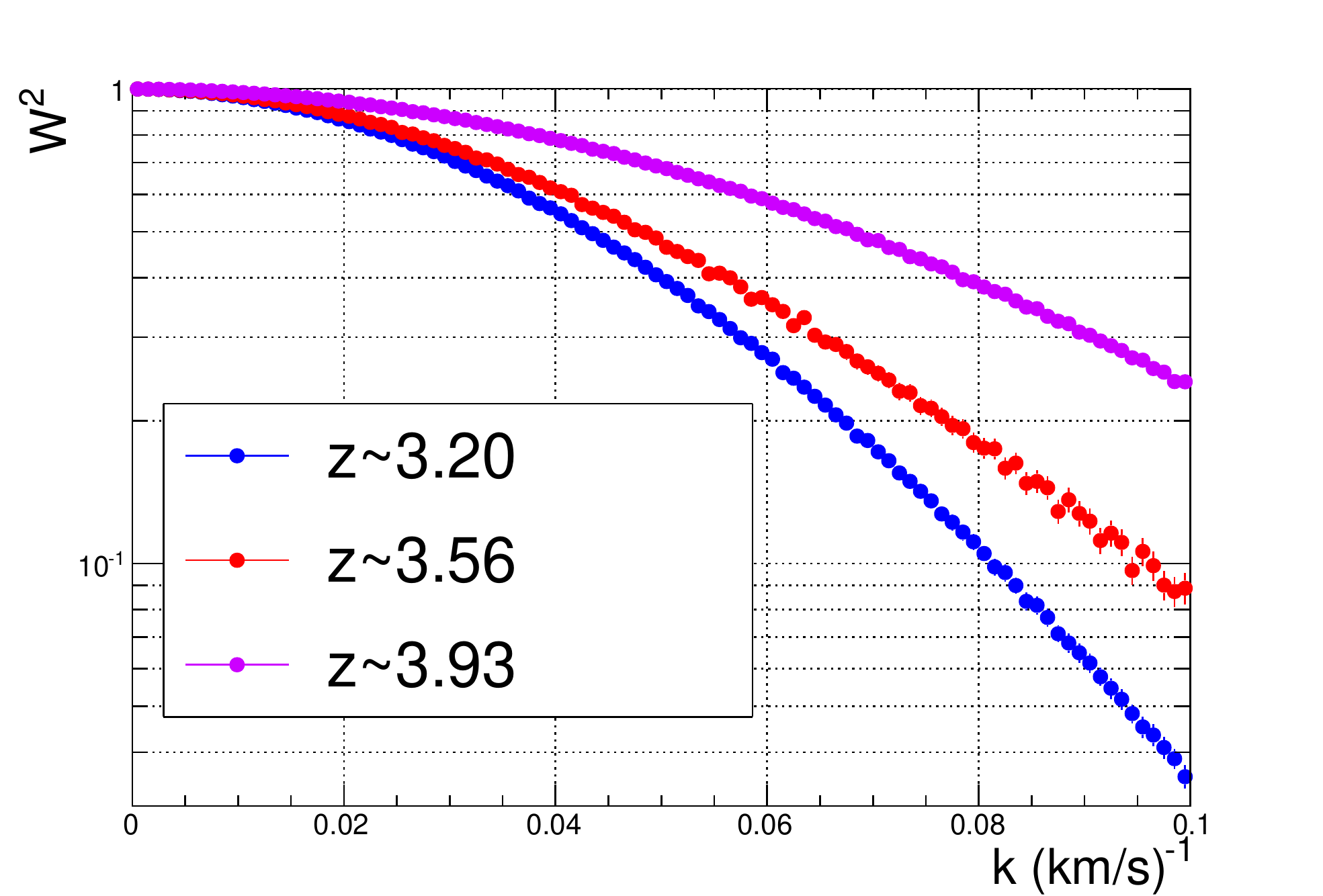,width = 7.5cm}
\epsfig{figure= 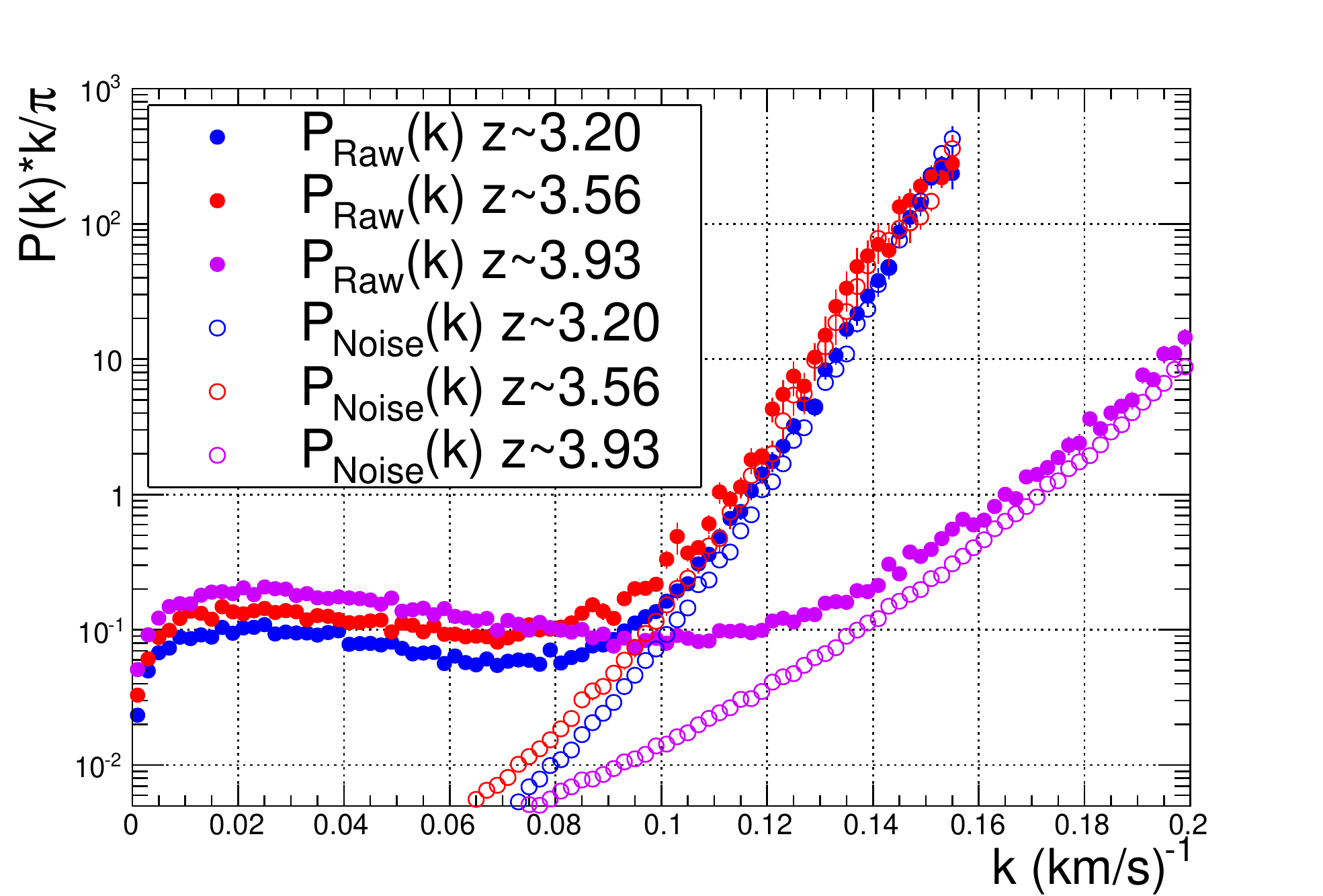,width = 7.5cm}
\caption{\it  Left plot: Average window function $W^2(k,\overline{R},\Delta v)$ for each of the three redshift bins $z=3.20$, 3.56 and 3.93 considered in this analysis. Right plot: $P^{raw}(k)$ (large dots) and $P^{noise}(k)$ (open circles) for the same redshift bins.}
\label{fig:ResoNoise}
\end{center}
\end{figure}

The window function $W^2(k,R,\Delta v)$ corresponding to the spectral response of the spectrograph depends on the two parameters
 $\Delta v$ and $R$ which are respectively the  pixel width and the spectrograph resolution:
 \begin{equation*}
 W(k,R,\Delta v)= \exp\left(- \frac{1}{2}(kR)^2\right) \times \frac{\sin(k\Delta v /2)}{(k\Delta v/2)}\,.
\end{equation*}
 The determination of $R$ is discussed in Sec.~\ref{sec:reso}. We illustrate in Fig.~\ref{fig:ResoNoise} (left) the  impact of the spectrograph resolution on the window function $W^2(k,R,\Delta v)$ for the various redshift bins considered in this work. 

The noise power spectrum $P^{noise}(k,z) $ is computed as a white noise, using, for each pixel, the photometric error  from the SDP spectra.   Fig.~\ref{fig:ResoNoise} (right) shows  $P^{raw}(k)$  and $P^{noise}(k)$ as a function of $k$. At small scales, i.e., high $k$ values, where $P^{noise}(k,z) $ is  dominant, the raw power spectrum $P^{raw}(k)$ asymptotically approaches  our estimate of $P^{noise}(k,z) $, thus providing a clear validation of our noise model. Because of the uncertainty on the resolution correction, we  limit our study to the following upper bounds in $k$:  $ 0.050, 0.060$ and $0.070\,{\rm s\,km^{-1} }$   for the three redshift bins $z=3.20$, 3.56 and 3.93, respectively. As a consequence, $P^{noise}(k,z) $ is always a few orders of magnitude smaller than the Ly$\alpha$ power spectrum and thus has a limited impact on the Ly$\alpha$ power spectrum measurement.

\subsubsection{Correction of the metal absorption}

The cross-correlated background  due to correlated absorption by Ly$\alpha$ and \ion{Si}{iii} within the Ly$\alpha$ forest can be estimated directly in the power spectrum. Since \ion{Si}{iii} absorbs at $\lambda = 1206.50\,$\AA, just 9\,\AA\ away from Ly$\alpha$, it appears in the power spectrum as wiggles with a frequency corresponding to  $\Delta v \sim 2271 \, {\rm km\,s^{-1}}$ . These oscillations are clearly seen on the BOSS power spectrum as shown on Fig.\ref{fig:P1D_BOSS_with_fit}. Their contribution cannot be isolated from the Ly$\alpha$ power spectrum and is therefore  included in the model of Ly$\alpha$ power spectrum fitted to the data as done in PY15 and BP16. In a similar way, we  considered a possible additional correlated absorption with  \ion{Si}{ii} with a frequency of  $\Delta v \sim 5577 \, {\rm km\,s^{-1}}$. In PY15 and BP16 papers, we have not significantly detected such a correlation with \ion{Si}{ii}.

The uncorrelated background due to metal absorption in the Ly$\alpha$ forest is independent of   Ly$\alpha$ absorption and it cannot be estimated directly from the power spectrum measured in the Ly$\alpha$ forest. In~\cite{Palanque-Delabrouille2013} we  addressed this issue by estimating the background components in side bands located at longer wavelengths than the  Ly$\alpha$ forest region. We apply the same technique here. We measure the power spectrum in side bands and  subtract it from the Ly$\alpha$ power spectrum measured in the same gas redshift range. This method is purely statistical: for a given redshift bin, we use different quasars to compute  the Ly$\alpha$ forest and the metal power spectra. It cannot be applied to the lower redshift bins of this analysis, since there are then no lower-redshift quasars from which to compute the metal contribution. Therefore, we assume that the metal power spectrum is identical for all the redshift bins, in agreement with what was shown in Fig.19 of~\cite{Palanque-Delabrouille2013}.  

\begin{figure}[htbp]
\begin{center}
\epsfig{figure= 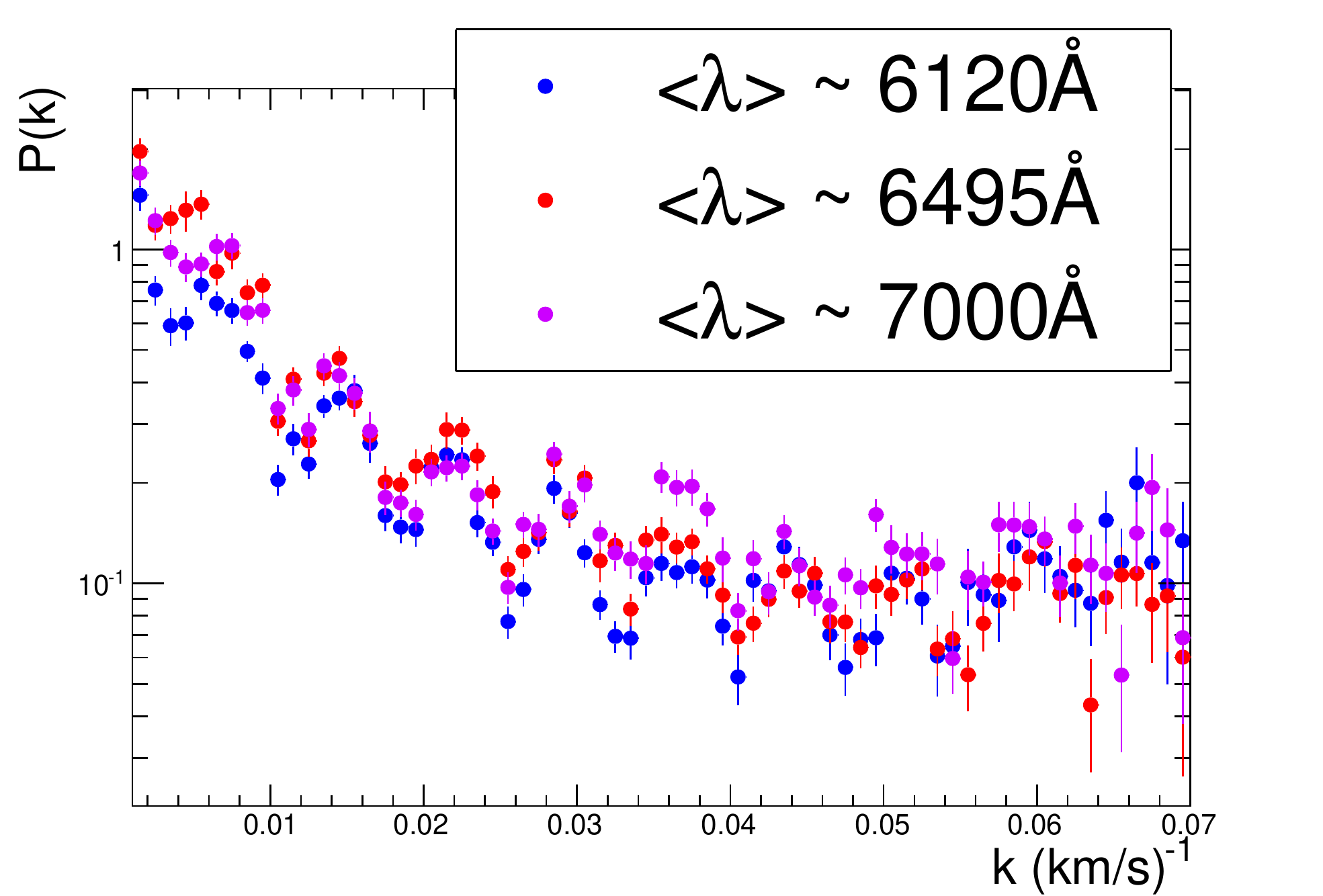,width = 7.5cm}
\epsfig{figure= 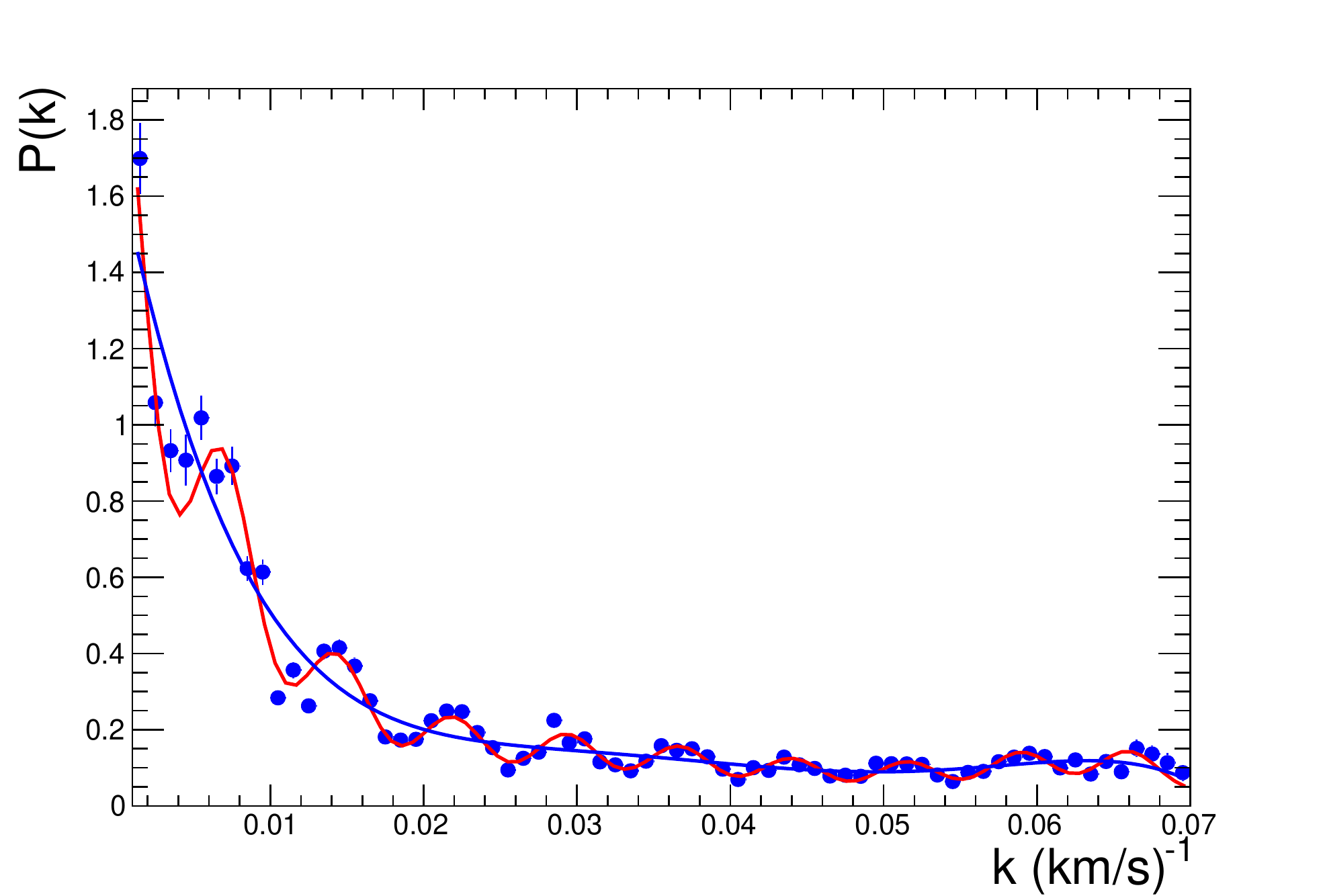,width = 7.5cm}
\caption{\it  Measurement of the  metal power spectrum $P(k)$ in the rest frame region $1270 - 1380\, \AA$. Left plot: $P(k)$  for three wavelength bins,  respectively centered on $6120$, $6495 $ and $7000\, \AA$. Right plot: instrumental-induced oscillation with a frequency $\Delta v = 855\; {\rm km\,s^{-1}}$  fitted on top of  the model for the metal power spectrum (red curve). The metal-only component  is given by the blue curve.} 
\label{fig:P1DMetals}
\end{center}
\end{figure}

In practice, we define one side band corresponding to the wavelength range  $ 1270 < \lambda_{\rm RF}< 1380 \,$\AA\  in the quasar rest frame. The  power spectrum measured in this side band  includes the contribution from all metals with  $\lambda_{\rm RF}>1380\,$\AA, including in particular absorption from \ion{Si}{iv} and \ion{C}{iv}. As shown in  Fig.~\ref{fig:P1DMetals} (left),   the amplitude of the effect is quite comparable for three different wavelength regions, demonstrating again that we can compute a unique estimate for  all redshift bins. However, we observe wiggles with a frequency corresponding to  $\Delta v \sim 855\, {\rm km\,s^{-1}}$ or $\Delta \lambda \sim 20\, \AA$. As this frequency varies with the tested wavelength region, and because it does not match  any doublet from expected metal emission lines, its origin is likely to be  instrumental. It may be a noise variation coming from the order structure of the echelle spectrum. The noise indeed increases and decreases in a periodic way due to the superposition of  adjacent orders, on a scale close to $ 20\; \AA$. As a consequence, we estimate the metal power spectrum from the wiggle-free broadband shape (blue curve on Fig.~\ref{fig:P1DMetals}, right).

\subsubsection{Estimator of $P^{{\rm Ly}\alpha}(k)$}

We determine the 1D power spectrum,  for three  bins of mean redshift 3.20, 3.56 and 3.93. We compute the Fourier transform using the  efficient FFTW package from~\cite{Frigo2012}.  The computation is done separately on each $z$-sector instead of  on the entire Ly$\alpha$ forest. The mean redshift of the Ly$\alpha$ absorbers in a  $z$-sector determines the redshift bin to which the $z$-sector contributes. We rebin the final power spectrum onto an evenly spaced grid in $k$-space, with $\Delta k = 0.001\,{\rm  s\,km}^{-1} $, giving equal weight to the different Fourier modes that enter each bin. Deriving Eq.~\ref{eq:P1D_raw}, the final 1D power spectrum, $P_{1D}(k)$ is obtained by averaging the corrected power spectra of all  contributing $z$-sectors from all quasars, as expressed in the following estimator of  $P^{{\rm Ly}\alpha}(k)$:  

\begin{equation}
P_{1D}(k)  =  \left< \frac{ P^{raw}(k) - P^{noise}(k) }{W^2(k,R,\Delta v)} \right> \,  -  P^{metals}(k),
\label{eq:P1D_FFT}
\end{equation}
where  $\langle\rangle$ denotes the ensemble average over quasar spectra.

\begin{figure}[htbp]
\begin{center}
\epsfig{figure= 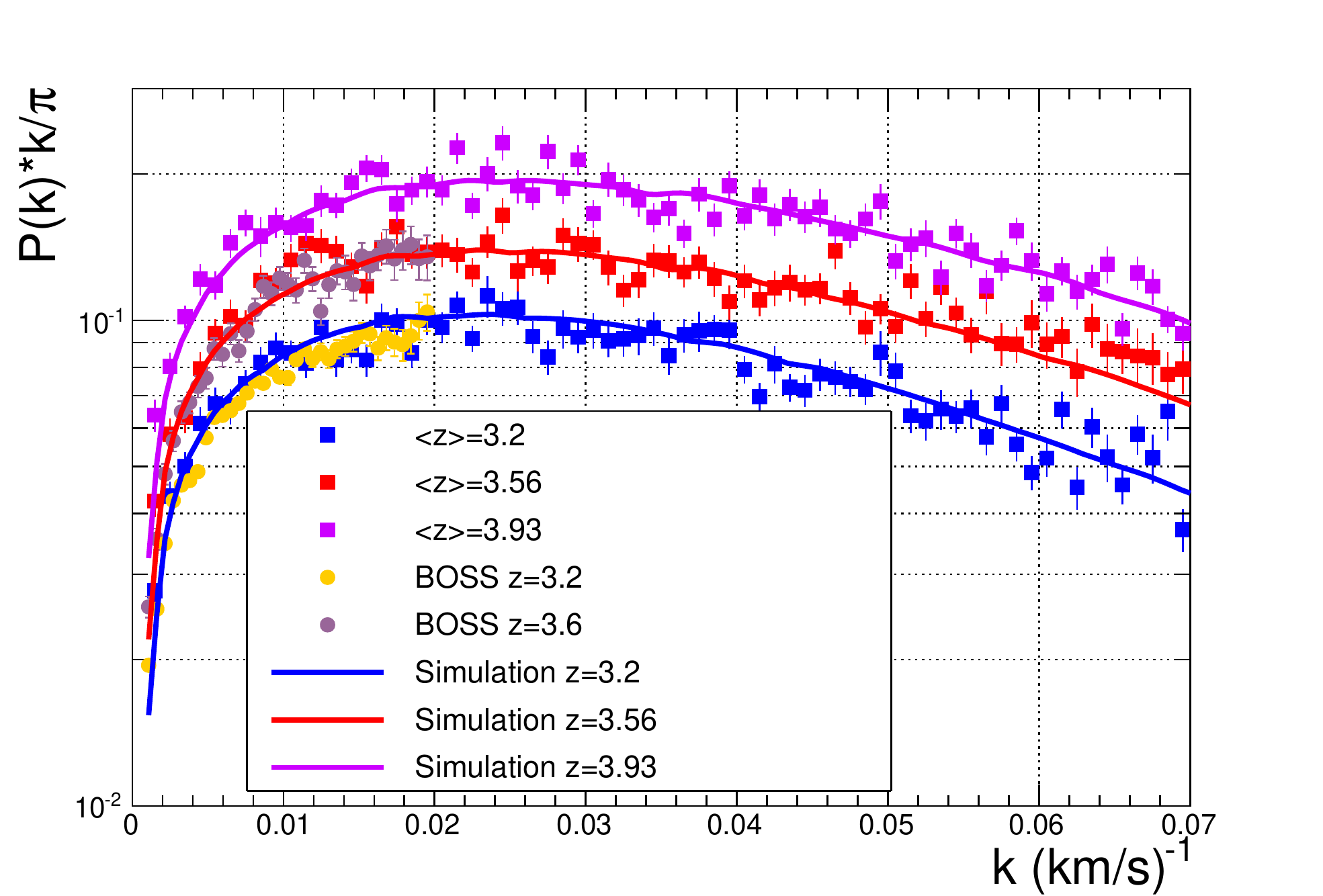,width = 7.5cm}
\epsfig{figure= 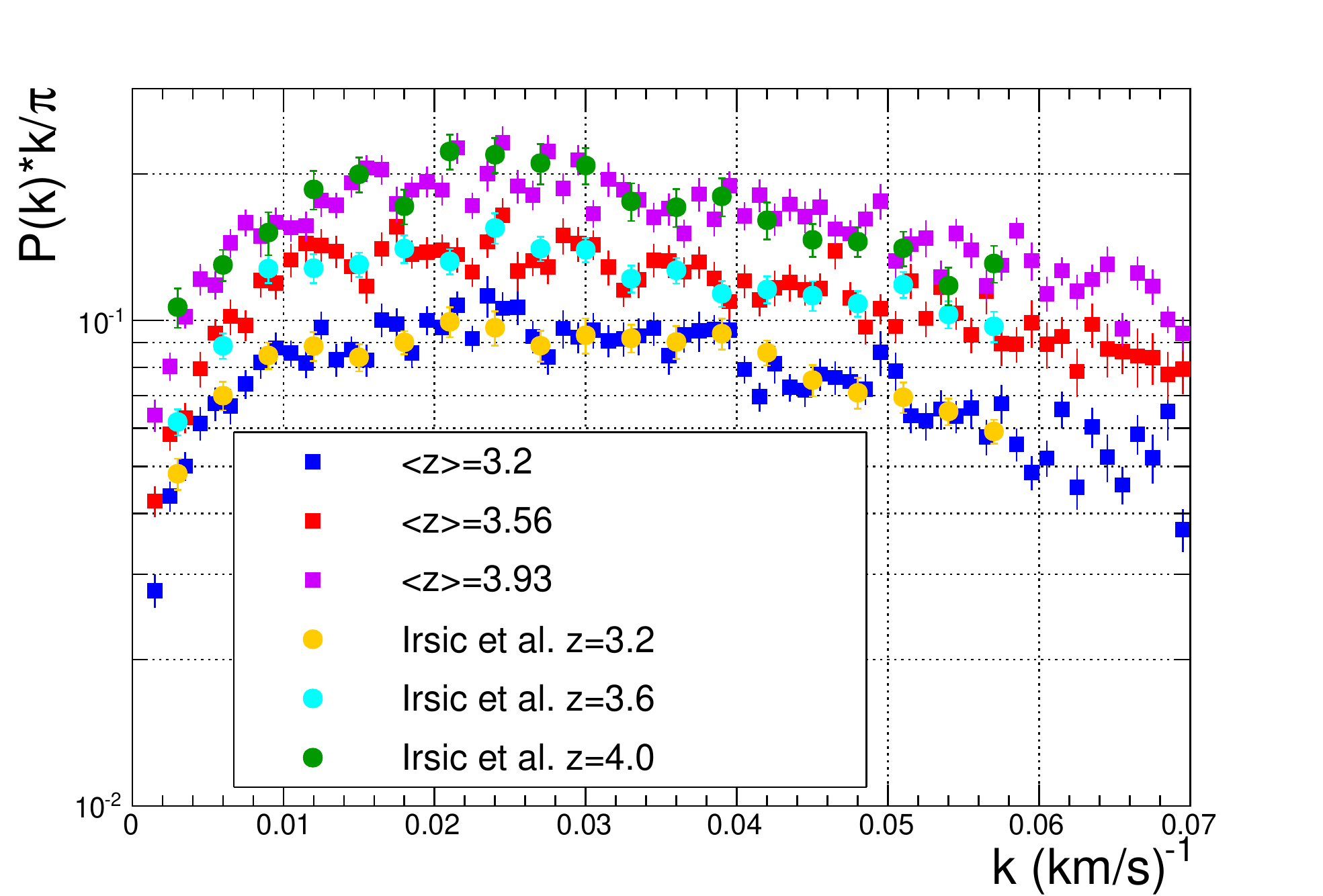,width = 7.5cm}
\caption{\it  Left plot: 1D Ly$\alpha$ forest power spectrum from XQ-100 spectra computed as explained in this paper and compared to BOSS power spectrum published in~\cite{Palanque-Delabrouille2013}. The  curves show the power spectrum derived from the simulations of~\cite{Borde2014}, corresponding to the "best guess" configuration ($\sigma_8=0.83$, $n_s=0.96$, and $\Omega_m=0.31$). The values of the power spectra are available in the appendix in Tab.~\ref{tab:Pk}. Right plot:  Same 1D Ly$\alpha$ forest power spectrum from  XQ-100 data, here  compared to the power spectrum computed in~\cite{Irsic2016} for the same data and three similar redshift bins. } 
\label{fig:P1D_XQ100}
\end{center}
\end{figure}

Figure~\ref{fig:P1D_XQ100} (left) shows the 1D Ly$\alpha$ forest power spectrum obtained with XQ-100 after subtraction of  the power spectrum of the metals. It  shows a good agreement with  previous BOSS measurements, although the spectrographs and the data are quite different.  The agreement with the other analysis of XQ-100~\cite{Irsic2016} is also remarkable, as is illustrated in Fig.~\ref{fig:P1D_XQ100} (right). The values $P_{1D}(k)$, the statistical uncertainty on $P_{1D}(k)$ and the noise power spectrum $P^{noise}(k,z) $ for the three redshift bins are given in the appendix in Tab.~\ref{tab:Pk}. In this study, we have assumed that the estimated errors are uncorrelated.

\section{Combining data}
\label{sec:data}

In this section, we first briefly introduce  the  data sets that we use to constrain the sum of neutrino masses and the mass of WDM. We then present  the hydrodynamical simulations we ran to interpret the 1D Ly$\alpha$ forest power spectrum, and    we explain the methodology followed in this paper,  which builds upon the one we developed for PY15 and BP16. 

\subsection{Data}

\subsubsection{\texorpdfstring{$\rm{Ly}\alpha$}{Lya}}
\begin{figure}[htbp]
\begin{center}
\epsfig{figure= 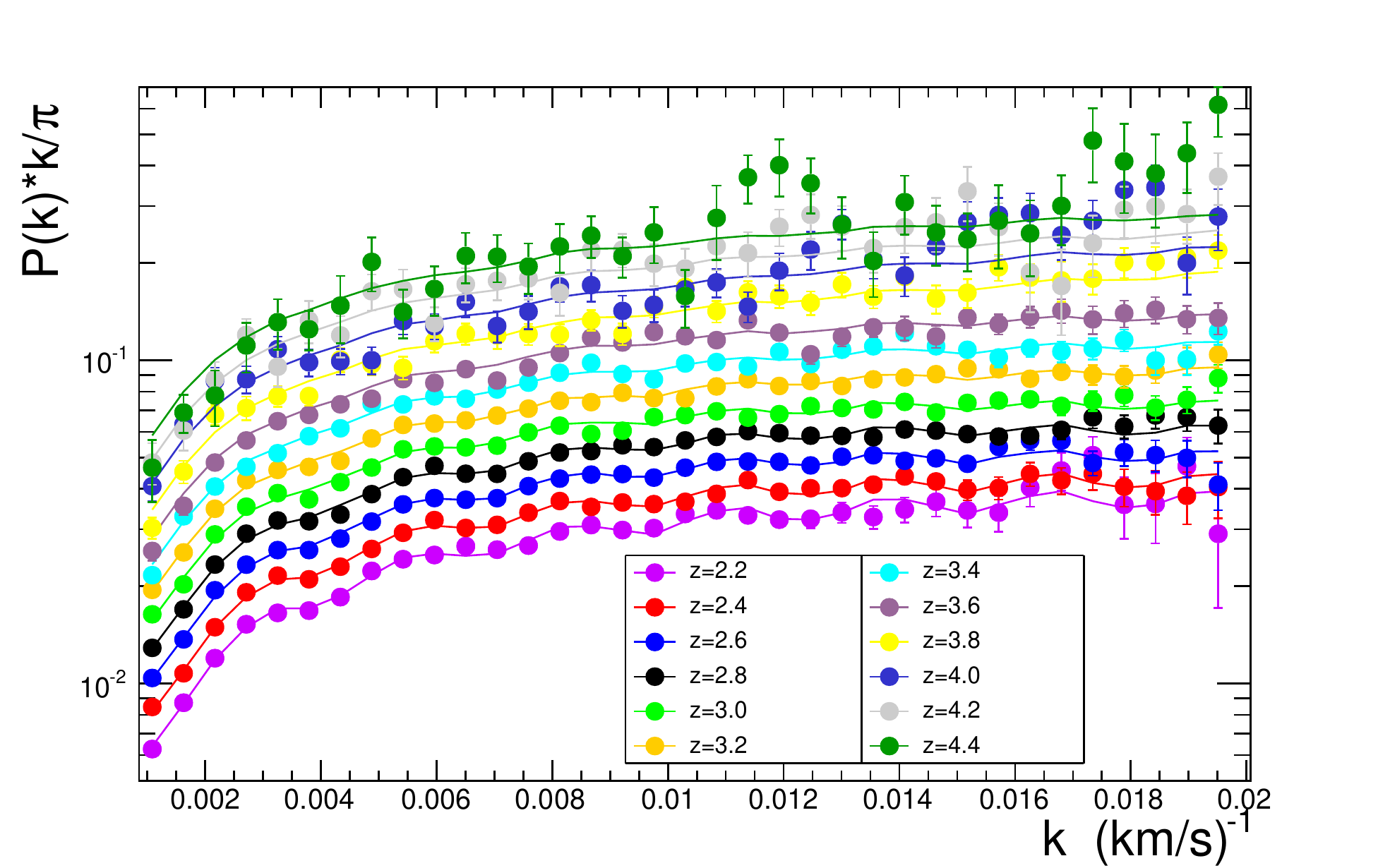,width = .8\linewidth}

\caption{\it 1D Ly$\alpha$ forest power spectrum from the SDSS-III/BOSS DR9 data.   The solid curves show the best-fit model when considering Ly$\alpha$ data alone.  The oscillations  arise  from Ly$\alpha$-Si\,III correlations, which occur at a  wavelength separation $\Delta \lambda = 9.2\,$\AA.} 
\label{fig:P1D_BOSS_with_fit}
\end{center}
\end{figure}

As our large-scale structure probe, we use the 1D Ly$\alpha$-flux power spectrum measurement from the first release of BOSS quasar data~\cite{Palanque-Delabrouille2013}. The data consist of a sample of $13~821$  spectra selected from the larger sample of about $60~000$ quasar spectra of the SDSS-III/BOSS DR9 \cite{Ahn2012, Dawson2012, Eisenstein2011, Gunn2006, Ross2012,Smee2013} on the basis of their high quality, high signal-to-noise ratio ($>2$) and good spectral resolution ($<85\,\rm km~s^{-1}$ on average over a quasar forest).  We  do the analysis on 420 Ly$\alpha$ data points, consisting of 12 redshift bins over $2.1<z<4.5$ and 35 $k$ bins with $k\le 0.020 \,{\rm  km\,s^{-1} }$ as shown on Fig.~\ref{fig:P1D_BOSS_with_fit}. 

\begin{figure}[htbp]
\begin{center}
\epsfig{figure= 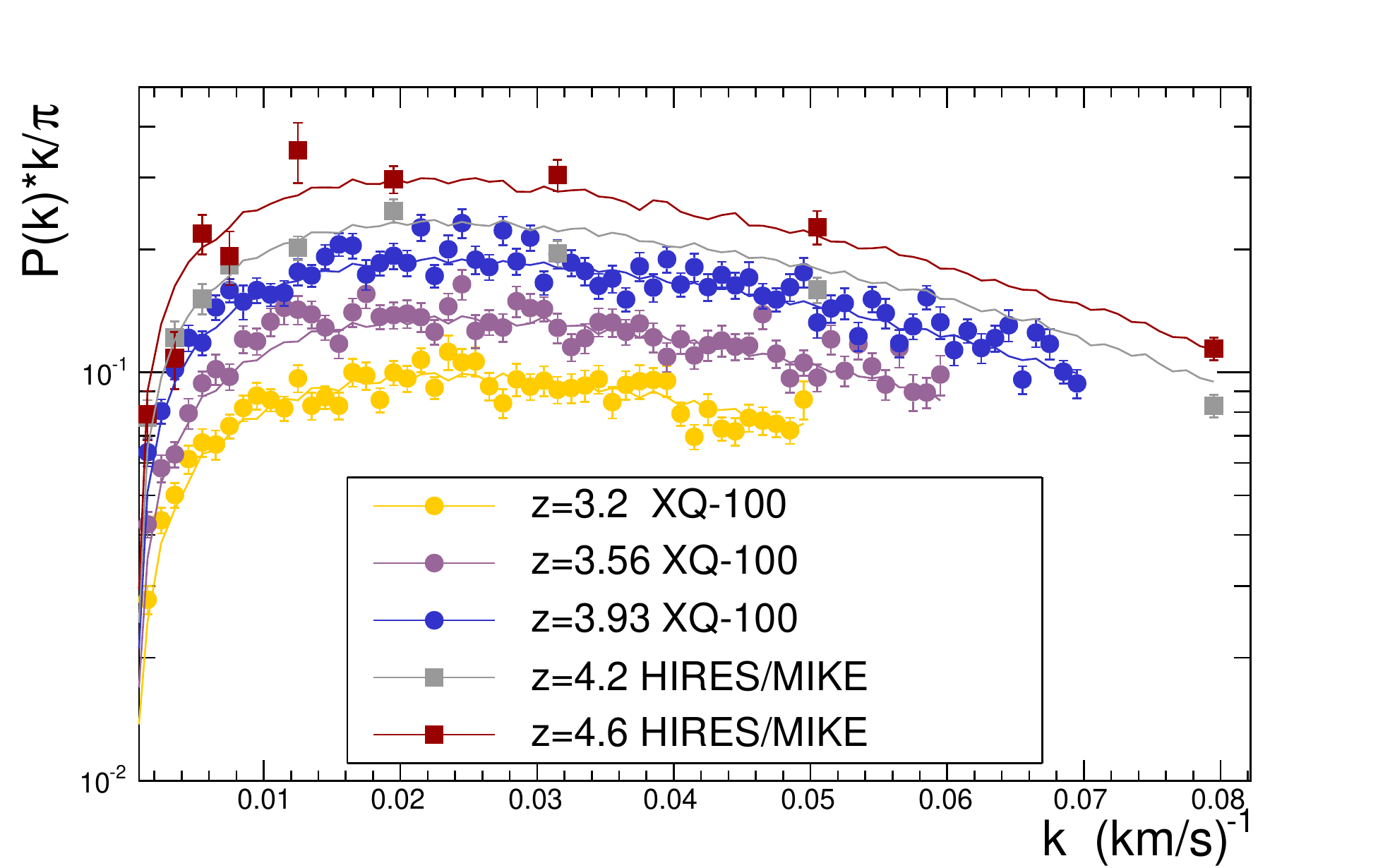,width = 7.5cm}
\epsfig{figure= 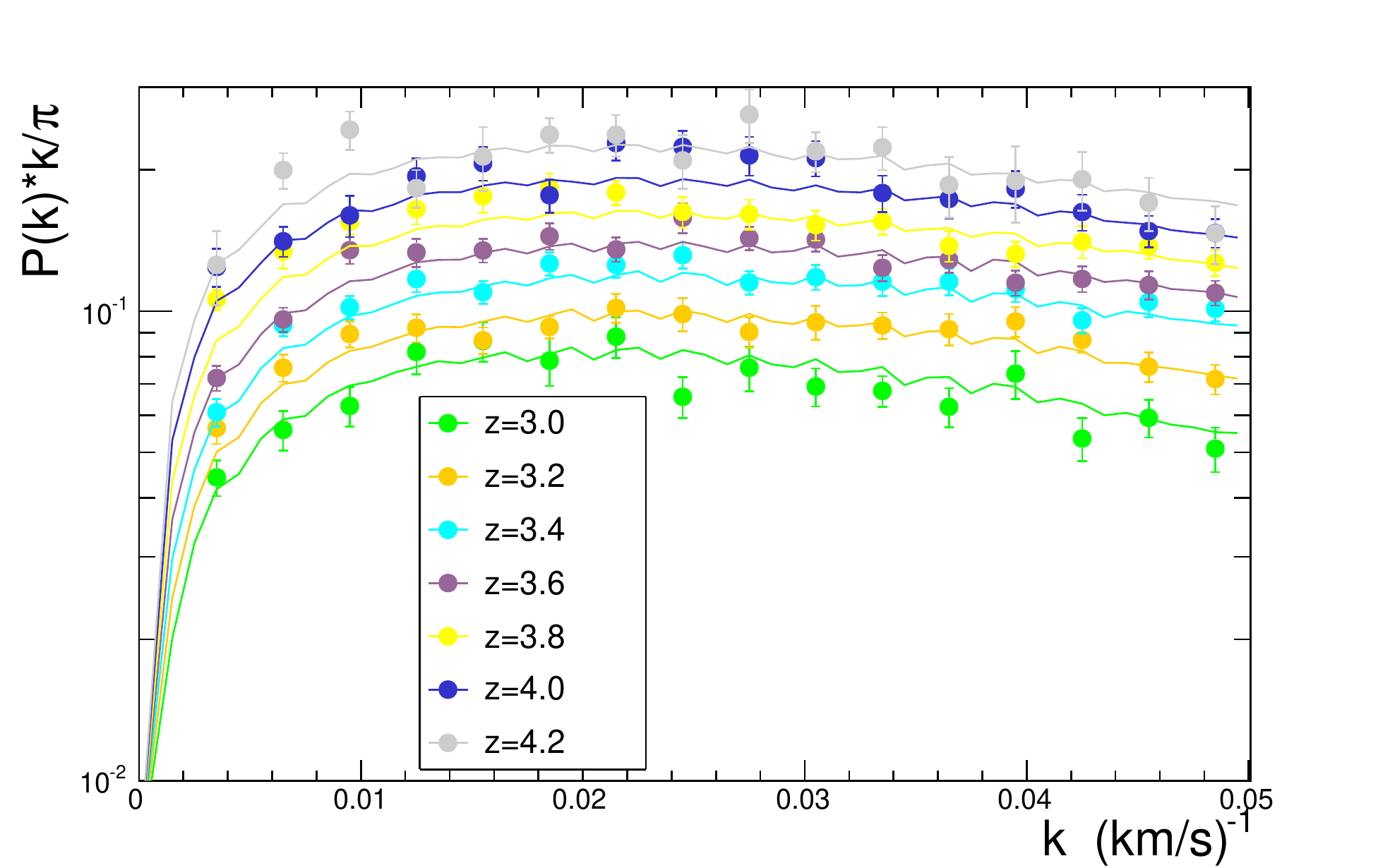,width = 7.5cm}
\caption{\it 1D Ly$\alpha$ forest power spectrum from XQ-100 data and HIRES/MIKE data.   The solid curves show the best-fit model when considering Ly$\alpha$ data from BOSS and XQ-100.  The oscillations  arise  from Ly$\alpha$-Si\,III correlations, which occur at a  wavelength separation $\Delta \lambda = 9.2\,$\AA\ and are driven by the BOSS data. The left plot corresponds to the analysis of the power spectrum presented in this paper. The right plot is the power spectrum measured in~\cite{Irsic2016}.}
\label{fig:P1D_XQ100_with_fit}
\end{center}
\end{figure}

We complement the BOSS 1D Ly$\alpha$-flux power spectrum with the XQ-100~\cite{Lopez2016} power spectrum. First, we use the power spectrum computed in Sec.~\ref{sec:xq100}, consisting of three redshift bins  at $z=3.2$, 3.56 and 3.93, and, respectively, 50, 60 and 70 $k$ bins corresponding to $k\le 0.050$, $0.060$ and $0.070\, {\rm  s\,km^{-1} }$  as shown in Fig.~\ref{fig:P1D_XQ100_with_fit} (left). Then, for the study of WDM in Sec~\ref{sec:WDM}, we also include in our likelihood the XQ-100 power spectrum measured by~\cite{Irsic2016}. This data set, shown in  Fig.~\ref{fig:P1D_XQ100_with_fit} (right), consists of 7 redshift bins with $2.9<z<4.3$ and 19 $k$ bins with $k\le 0.057\, {\rm  s\,km^{-1} }$.  

Finally, to the BOSS+XQ-100 data set, we further add the power spectrum measured by~\cite{Viel2013} with the high-resolution HIRES/MIKE spectrographs. We use  the lowest two redshift bins, $z=4.2$ and $z=4.6$. The power spectrum shown  in Fig.~\ref{fig:P1D_XQ100_with_fit} (left) was  obtained by combining in quadrature the HIRES and MIKE for  9 $k$ bins with $k\le 0.080\, {\rm  s\,km^{-1} }$.  The highest-redshift snapshot that we extracted from  our simulations (see Sec.~\ref{sec:simu}) is at $z=4.6$, which prevents us from using the other two redshift bins ($z=5.0$ and $z=5.4$) of the HIRES/MIKE data.

\subsubsection{Cosmic microwave background}\label{sec:cmb}
As in PY15, we use  cosmic microwave background (CMB) data and results  from the full Planck mission~ \cite{Planck2015}. We consider several subsets of Planck data. The base configuration, denoted `TT+lowP' as in  \cite{Planck2015}, uses the TT spectra at low and high multipoles and the polarization information up to multipoles $\ell=29$ (`lowP'). We also use at times the configuration based on TT, TE and EE spectra, along with the low-multipole polarization, denoted `TT+TE+EE+lowP'.  

\subsubsection{Baryon acoustic oscillations}
We  occasionally combine CMB data with measurements of the BAO scale by 6dFGS~\cite{Beutler2011}, SDSS main galaxy sample~\cite{Ross2014}, BOSS-LOWZ~\cite{Anderson2014} and CMASS-DR11~\cite{Anderson2014}. Theses measurements are henceforth globally denoted 'BAO'. The additional constraints that these measurement provide on cosmological parameters are included in the present work with their full correlation with CMB data. 
Both CMB and BAO constraints are taken from  the  Markov Chains publicly available through the official Planck Legacy Archive at http://pla.esac.esa.int. 

\subsection{Simulations}
\label{sec:simu}
To predict the Ly$\alpha$ flux power spectrum, we use the set of simulations extensively described in PY15 and BP16. The simulations are run using a  parallel tree smoothed particle hydrodynamics  (tree-SPH) code {\sc Gadget-3}, an updated version of the public code  {\sc Gadget-2} \cite{Springel2001,Springel2005}. The simulations are started at $z=30$, with initial transfer functions and power spectra computed with {\sc CAMB}~\citep{Lewis2000}, and initial particle displacements generated with second-order Lagrangian Perturbation Theory {\sc 2LPT}\footnote{\url{http://cosmo.nyu.edu/roman/2LPT/}}.  We include three particle types: collisionless dark matter, gas, and, when relevant, degenerate-mass neutrinos.  We showed in \cite{Palanque-Delabrouille2015} that considering inverted or normal neutrino mass hierarchy  yields a flux power spectrum that differs by less than 0.05\%  from the degenerate-mass scenario, a level ten times lower than the statistical uncertainties in the simulation and almost two orders of magnitude below the data uncertainties. The degenerate-mass hypothesis is thus highly justified.  The simulations cover the volume of a periodic 100~Mpc$/h$ box containing the equivalent of $3072^3$ particles of each type.  Following a method originally suggested in \cite{McDonald2003}, we obtain this resolution by splicing together large-volume  and  high-resolution simulations,  using a transition simulation that corrects the large box for its lack of coupling between small and large modes, and the high-resolution simulation for its small volume. We studied the accuracy of the splicing technique in \cite{Palanque-Delabrouille2015} and we leave parameters  free when fitting the data to account for residual biases. 

The cosmological parameters are centered on the Planck 2013 best-fit values  \cite{PlanckCollaboration2013}. Using simulations where one or two parameters at a time are given off-centered values, we compute first and second-order derivatives of the Ly$\alpha$ flux power spectrum with respect to each parameter,  which we use to derive a second-order Taylor expansion of the predicted Ly$\alpha$ flux power spectrum. The cosmological parameters cover the range $H_0=67.5\pm 5~{\rm km\,s^{-1}\,Mpc^{-1}}$, $\Omega_M=0.31\pm0.05$, $n_s=0.96\pm0.05$,  $\sigma_8=0.83\pm0.05$. In all the runs, we keep $\Omega_b = 0.0221$. While our central simulation assumes massless neutrinos, some runs include neutrinos with masses $\sum m_\nu=0.4$ or 0.8~eV. Where WDM is assumed,  the dark matter particles are thermal relics with masses $m_X=2.5$ or 5.0~keV. All our simulation runs start at $z = 30$, with initial conditions having the same random seed. Snapshots are produced at regular intervals in redshift from $z = 4.6$ to 2.2, with $\Delta z = 0.2$.

We  test the influence of assumptions on the IGM astrophysics  by running simulations for central and offset values of  relevant parameters. The photo-ionization rate of each simulation is fixed by requiring the effective optical depth at each redshift to follow the empirical law $\tau_{\rm eff}(z) =  A^\tau  (1+z) ^{\eta^\tau}$, with 
$A^\tau=0.0025 \pm 0.0020$ and $\eta^\tau=3.7\pm0.4$~in agreement with~\citep{Meiksin2009}. This renormalization is done at the post-processing stage, as justified in \cite{Theuns2005}, allowing us to test the impact of different scalings without running new simulations. The IGM temperature-density  relation $T = T_0 \Delta^{\gamma-1}$ is obtained using simulations ran for $\gamma(z=3)=1.3\pm 0.3$ and $T_0 (z=3) = 14000\pm 7000\,$K.   We use the quick-Ly$\alpha$ option to convert gas particles with overdensities exceeding $10^3$ and temperature below $10^5$~K into stars.

\subsection{Methodology}
\label{sec:method}
The analysis of the data is done with a frequentist approach. We showed in~\cite{Palanque2015a} that the constraints obtained with either a frequentist or a Bayesian (MCMC) approach were in excellent agreement. The large number of nuisance parameters that are now included  in the fit, however, prevents the use of a Bayesian method where convergence is then hard to reach. 

The parameters that are  floated in the minimization procedure belong to three categories. The first category models a flat $\Lambda$CDM cosmology with free $H_0$, $\Omega_M$, $n_s$, $\sigma_8$ and either $\sum m_\nu$, $m_X$ or $m_s$. The second category describes the IGM, letting free the parameters described in table~\ref{tab:astroparam}, namely $T_0$, $\gamma$, $\eta^{T_0}(z<3)$, $\eta^{T_0}(z>3)$, $\eta^{\gamma}$, $A^\tau$, $\eta^\tau$ and two amplitudes for the correlated absorption of Ly$\alpha$ with \ion{Si}{iii}  or \ion{Si}{ii}. In particular, to take into account the redshift evolution of $T_0(z)$, $\gamma(z)$, we modeled them using a single power law for $\gamma$ and a broken power law at $z=3$ for $T_0$ as explained in~\cite{Palanque-Delabrouille2015}. 

Finally, the last category groups all nuisance parameters that allow us to account for uncertainties or corrections related to noise in the data, spectrograph resolution, modeling of the IGM, residual bias in the slicing technique, supernova and AGN feedbacks, and redshift of reionization. Details on the fit parameters and on the dependance with scale and redshift of the nuisance parameters can be found in \cite{Palanque-Delabrouille2015}.


\section{{Constraints on   \texorpdfstring{$\Lambda$CDM$\nu$}{LCDMnu}}}
\label{sec:LambdaCDM}

In this section, we present the cosmological results of our analysis considering $\Lambda$CDM$\nu$ cosmology with standard active neutrinos. The analysis extends the method presented in PY15 to the XQ-100 data. We first describe the constraints obtained with Ly$\alpha$ alone and  we then add  CMB and BAO data.

\subsection{\texorpdfstring{$\Lambda$CDM$\nu$}{LCDMnu} cosmology  from  \texorpdfstring{$\rm{Ly}\alpha$}{Lya} data alone}
\label{sec:LyaAlone}

The results for Ly$\alpha$ alone are shown in columns (1) to  (3) of  Tab.~\ref{tab:fit_standard_nu}.  Column (1) recalls the results of PY15 for BOSS Ly$\alpha$ alone. The maximization of the likelihood with  Ly$\alpha$ data, imposing  a Gaussian constraint $H_0 = 67.3 \pm 1.0$,  gives a  best-fit value of  the sum of the neutrino masses $\sum m_\nu$ equal to 0.41~eV and  compatible with 0 at about $1\sigma$. The upper bound on $\sum m_\nu$ is thus $1.1$~eV (95\% C.L.).
   
In column (2) of Tab.~\ref{tab:fit_standard_nu},  we present the results  with  XQ-100 alone, using $P(k)$ measured with the method described in Sec.~\ref{sec:xq100} .  The upper bound on $\sum m_\nu$ is $1.2$~eV (95\% C.L.), compatible with the upper bound from BOSS alone. 
 
The other astrophysical and cosmological parameters are in very good agreement between BOSS-alone and XQ-100-alone. The largest differences observed are for $T_0$ and $\sigma_8$. They represent respectively $1.7\sigma$ and $1.4\sigma$. However, the comparison of BOSS and XQ-100 and the interpretation of differences in the best-fit parameters may be delicate as all the variables are correlated. In particular, we illustrate in Fig.~\ref{fig:AstroCosmoLyaPlanck} (left)  the strong anti-correlation between  $T_0$ and $\gamma$. As a result, a low value of $\gamma$  pushes $T_0$ to high values. In this neutrino-mass-oriented analysis, the likelihood is built in such a way as not to be too sensitive to underlying assumptions on  IGM parameters, which we here treat as nuisance parameters. The shapes of $T_0(z)$ and $\gamma(z)$ are let free in the maximization of the likelihood as explained in Sec.~\ref{sec:method}.

 Finally, by combining BOSS and  XQ-100, see column (3), the best-fit value  of $\sum m_\nu$ decreases to 0.34~eV and the upper bound on $\sum m_\nu$ tightens to $0.8$~eV (95\% C.L.).  The fitted values of  astrophysical and nuisance parameters are all well within the expected range.  The neutrino mass is correlated to  $\sigma_8$ (-26\%),  $n_s$ (19\%) and  $\Omega_m$ (33\%). Correlations between all other cosmological parameters have smaller amplitudes.

\begin{table}[htbp]

\caption{\it Best-fit value and 68\% confidence levels of the cosmological parameters of the model fitted to the flux power spectrum  measured with the BOSS Ly$\alpha$ data 	and XQ-100 Ly$\alpha$  data combined with several other data sets.  Column (1) shows the results of PY15 with BOSS alone~\cite{Palanque-Delabrouille2015}.  Column (2) shows the results with  XQ-100 alone, using $P(k)$ measured with the method described in Sec.~\ref{sec:xq100} . Column (3) shows the results for the combined fit of  BOSS and XQ-100. For columns (1-3), we used a Gaussian constraint, $H_0 = 67.3 \pm 1.0$. The following columns are obtained by combining Ly$\alpha$ with CMB and BAO.}

\begin{center}
{\small
\begin{tabular}{lccccc}
\hline
 &   Ly$\alpha$   &   Ly$\alpha$  &     Ly$\alpha$  &   Ly$\alpha$ &  Ly$\alpha$  \\
 &  BOSS & XQ-100 & BOSS + XQ-100 & BOSS + XQ-100 & BOSS + XQ-100 \\
Parameter &  + $H_{0}^\mathrm{Gaussian}$  &  + $H_{0}^\mathrm{Gaussian}$  & + $H_{0}^\mathrm{Gaussian}$
&   + Planck &  + Planck  \\
 &    &   &  & ({\scriptsize TT+lowP } ) &   ({\scriptsize TT+TE+EE+lowP}) \\
 &    &   &  & & + BAO   \\
 & (1) & (2)  & (3) & (4) & (5) \\
 \hline \\[-10pt]

$T_0$ (z=3) {\scriptsize($10^3$K)}  & $8.9\pm3.9$   & $21.4\pm6.0$    & $14.7\pm3.3$ & $16.1\pm2.5$  & $16.5\pm2.5$ \\[2pt]
$\gamma$   &  $0.9\pm0.2$   &  $0.88\pm0.5$      &  $1.0\pm0.2$& $0.7\pm0.2$ & $0.7\pm0.2$\\ [2pt]
$\sigma_8$ &  $0.831\pm0.031$    & $0.738 \pm 0.059$    & $0.783 \pm 0.023$ & $0.837\pm 0.017$ & $ 0.830\pm 0.015$\\[2pt]
$n_s$ &  $0.939\pm0.010$     &  $0.920\pm0.023$     &  $0.950\pm0.008$ &  $0.962 \pm 0.004$ & $0.961 \pm 0.004$\\[2pt]
$\Omega_m$  &  $0.293\pm0.014$     &  $0.317\pm0.024 $      &  $0.282\pm0.012 $ &  $0.288\pm0.013$& $0.310 \pm 0.007$\\[2pt]
$H_0$~{\scriptsize(${\rm km~s^{-1}~Mpc^{-1}}$)}   & $67.3\pm1.0$      &   $67.2 \pm 1.0$     &   $67.1 \pm 1.0$ & $69.1\pm 1.0$ & $67.7 \pm 0.6$ \\[2pt]
$\sum \! m_\nu$~{\scriptsize(eV)} &$<1.1 $~{\scriptsize(95\% CL)} & $<1.2$ ~{\scriptsize(95\% CL)} & $<0.8$ ~{\scriptsize(95\% CL)}& $< 0.14$~{\scriptsize(95\% CL)} & $<0.14$~{\scriptsize(95\% CL)} \\[2pt]
\hline
\end{tabular}
}
\end{center}

\label{tab:fit_standard_nu}
\end{table}

\subsection{\texorpdfstring{$\Lambda$CDM$\nu$}{LCDMnu} cosmology from  \texorpdfstring{$\rm{Ly}\alpha$}{Lya} data and other probes }
\label{sec:LyaCMB}

We now  combine the  Ly$\alpha$ likelihood (imposing no constraint on $H_0$) with the likelihood of Planck 2015 data that we derive from the central values and covariance matrices available in the official 2015 Planck repository. As in the previous section, we  focus on the base $\Lambda$CDM$\nu$ cosmology. Column (4) of Tab.~\ref{tab:fit_standard_nu} shows the results for the combined set of Ly$\alpha$ (BOSS and XQ-100) and the base configuration we chose for Planck data, i.e. (TT+lowP) (cf. details in Sec.~\ref{sec:data}). In  column (5),  we extend the CMB measurements to (TT+TE+EE+lowP) and we add BAO data. 

\begin{figure}[htbp]
\begin{center}
\epsfig{figure= 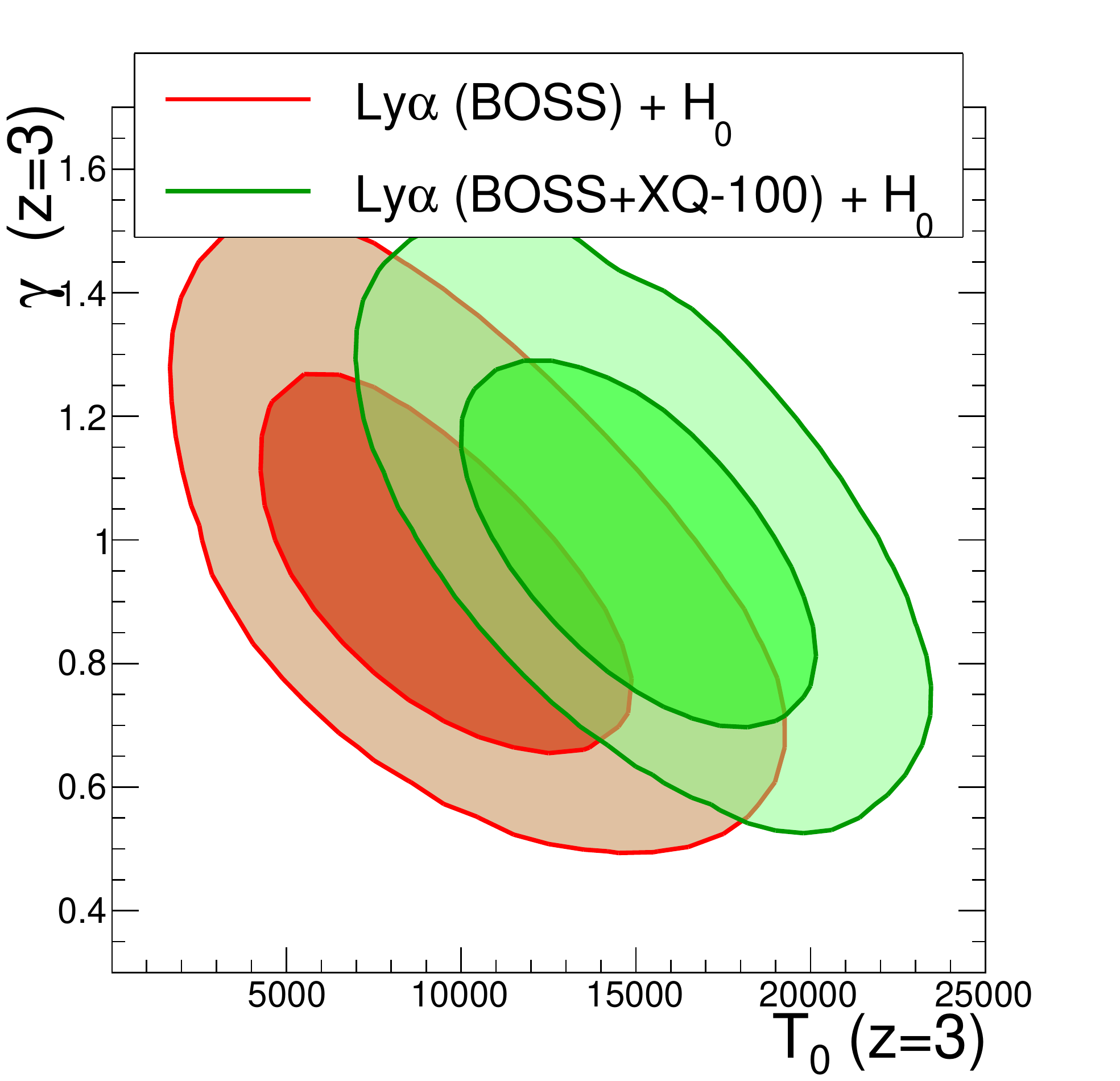,width = 7.5cm}
\epsfig{figure= 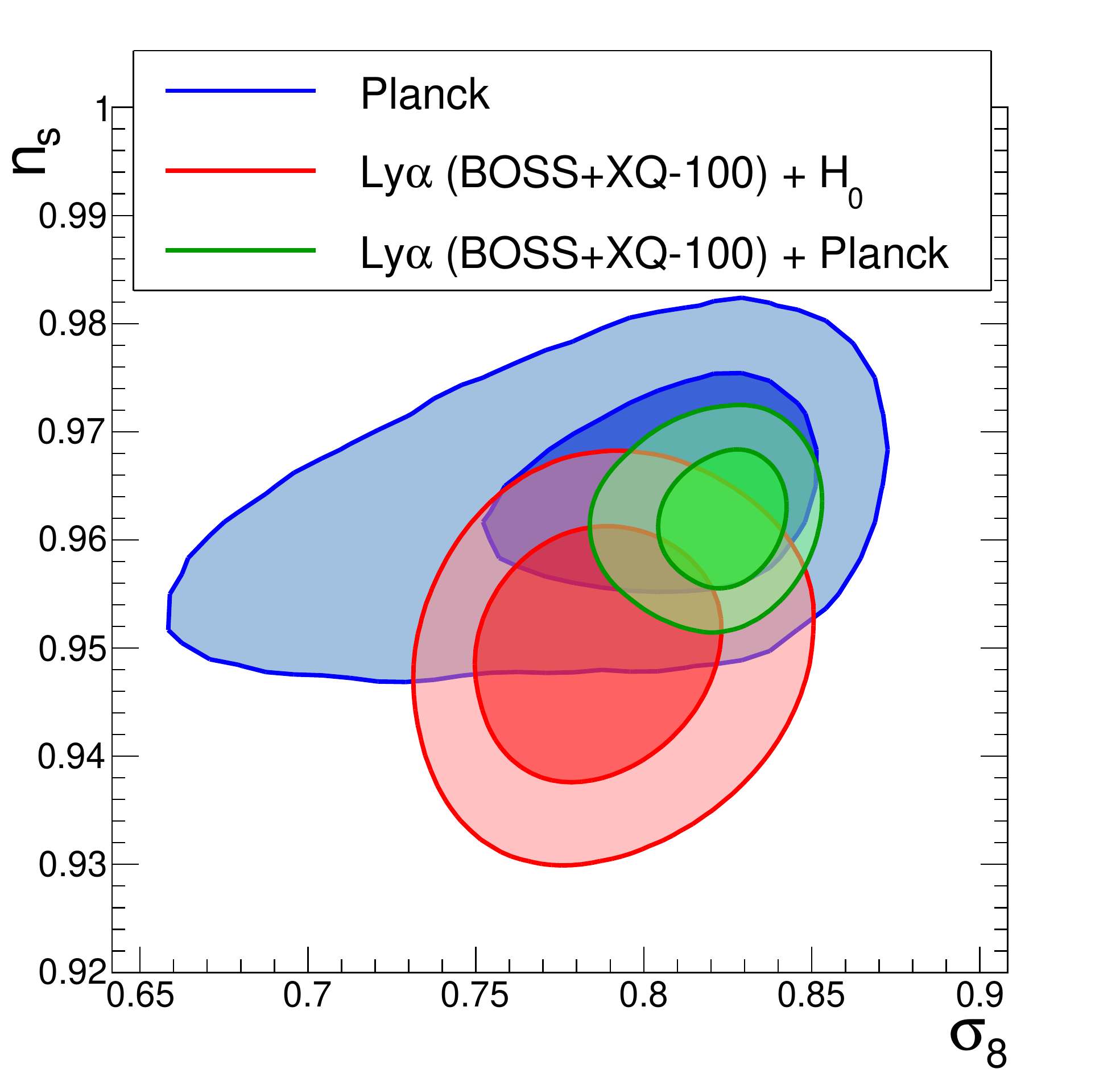,width = 7.5cm}
\caption{\it   2D confidence level contours for  the $(T_0,\gamma)$ astrophysical parameters defined at redshift $z=3$ and for the $(\sigma_8 , n_s)$    cosmological parameters.  Left Plot:   68\% and 95\% confidence contours  obtained for the BOSS Ly$\alpha$ data with a Gaussian constraint $H_0 = 67.3 \pm 1.0~{\rm km~s^{-1}~Mpc^{-1}}$ (red), and for the combination of BOSS and XQ-100 Ly$\alpha$ data (green).   Right plot:  68\% and 95\% confidence contours  obtained for the BOSS and XQ-100 Ly$\alpha$ data with a Gaussian constraint $H_0 = 67.3 \pm 1.0~{\rm km~s^{-1}~Mpc^{-1}}$ (red), for the Planck 2015  TT+lowP data (blue) and for the combination of  Ly$\alpha$ and Planck 2015 (green). }
\label{fig:AstroCosmoLyaPlanck}
\end{center}
\end{figure}

We illustrate in Fig.~\ref{fig:AstroCosmoLyaPlanck} (right) the main 2D contours on cosmological parameters.  The small  tension  on $n_s$ between Ly$\alpha$ and Planck data, observed in~\cite{Palanque2015a,Palanque-Delabrouille2015}, is still present when including XQ-100 data, although its significance is reduced to 1.7~$\sigma$. We  demonstrated  in PY15, anyhow, that the tension on $n_s$ has little effect on the constraint on $\sum m_\nu$ because  of the mild  correlation between these two parameters  (19\% in Ly$\alpha$, -45\% in Planck TT+lowP).  As was already seen in PY15, $\sum m_\nu$ is mostly correlated to $\sigma_8$ (-26\% in Ly$\alpha$, -95\% in Planck TT+lowP) and to $\Omega_m$ (19\% in Ly$\alpha$, 92\% in Planck TT+lowP).   

\begin{figure}[htbp]
\begin{center}
\epsfig{figure= 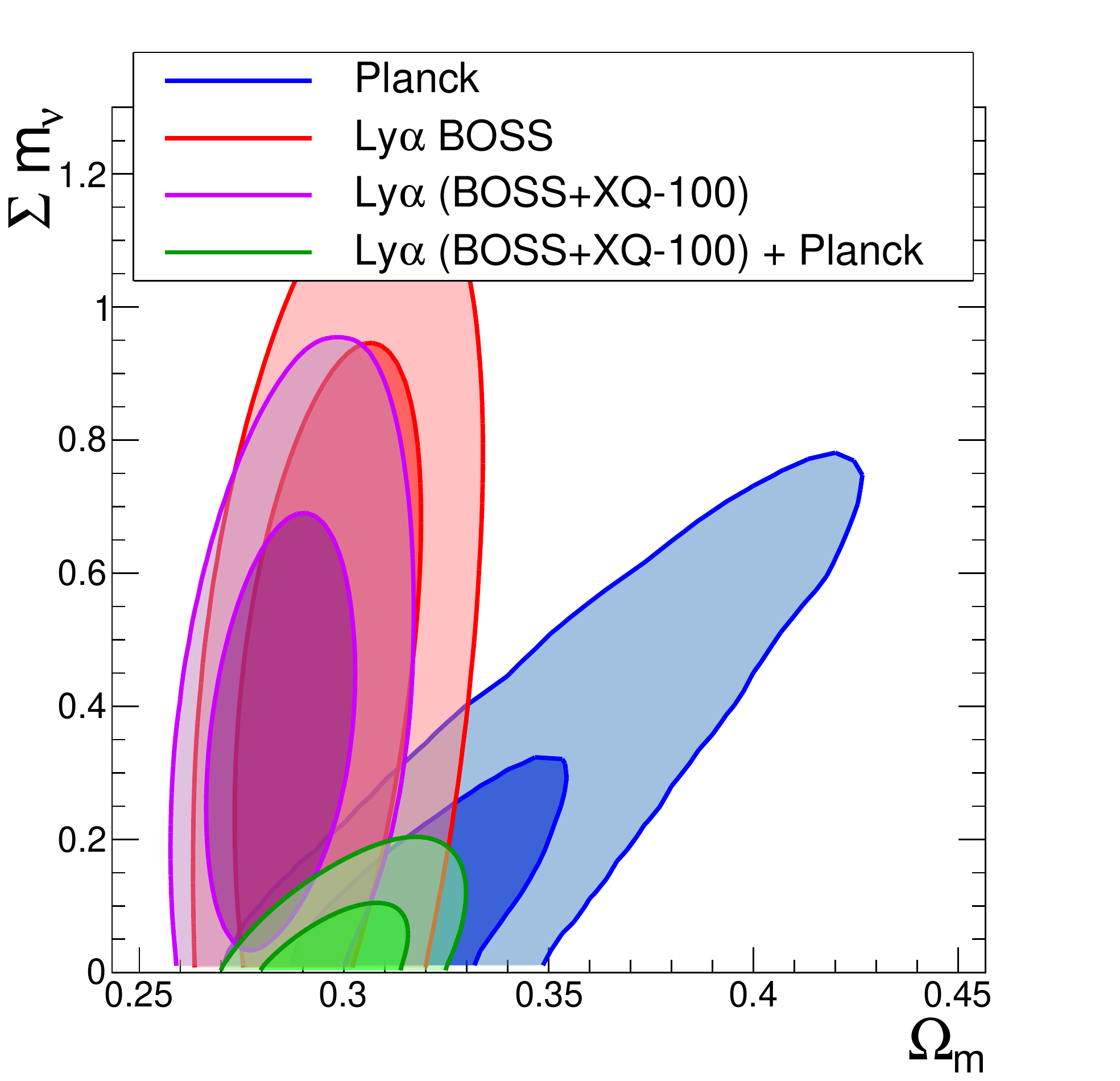,width = 7.5cm}
\epsfig{figure= 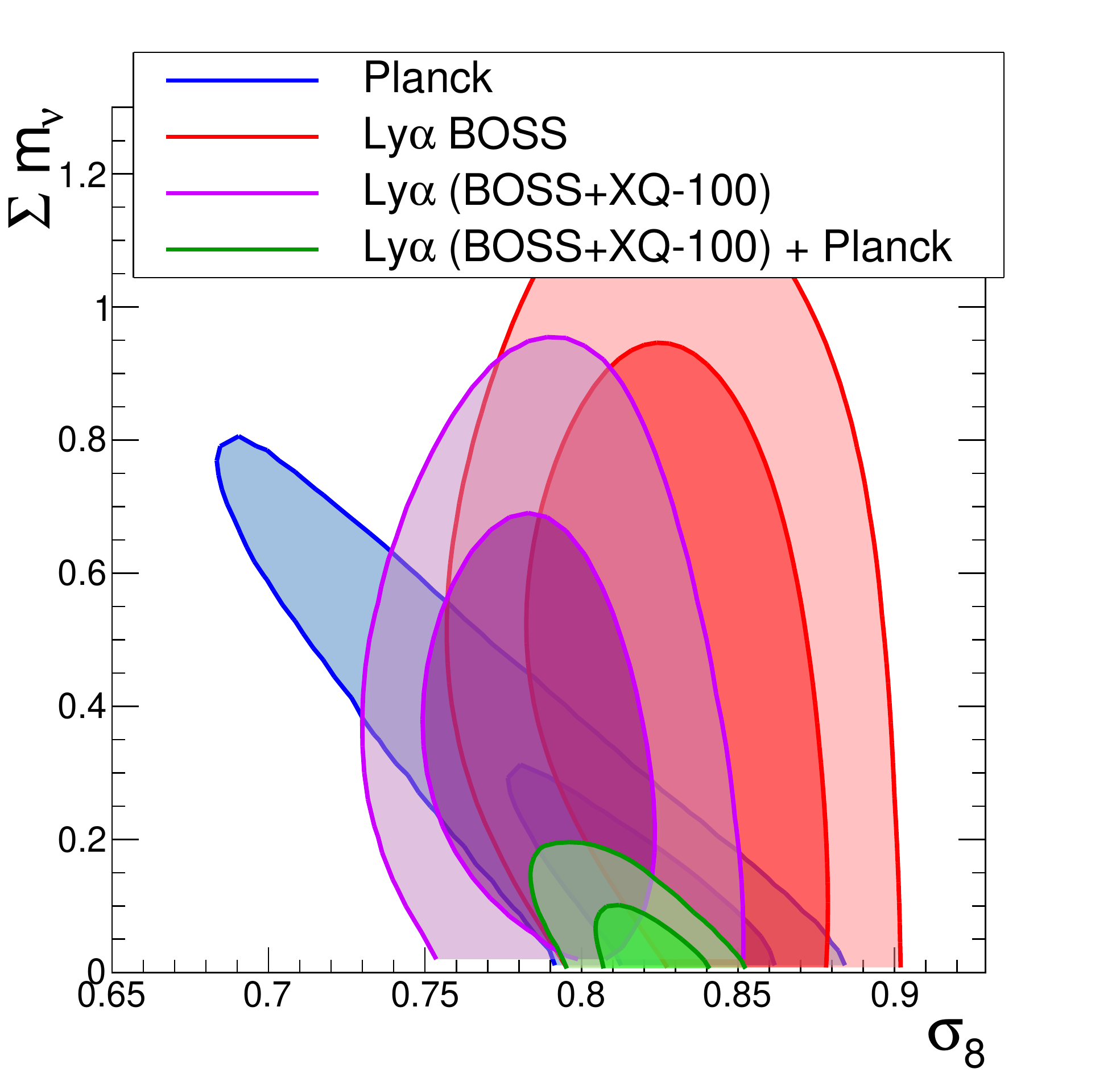,width = 7.5cm} 
\caption{\it    The 68\% and 95\% 2D confidence level contours for  Planck 2015  TT+lowP data (blue),  BOSS data with a Gaussian constraint on $H_0$ (red), then by adding XQ-100 data to BOSS data (magenta) and finally by combining BOSS, XQ-100 and Planck (green). Left plot: 2D confidence level contours for   $(\Omega_m , \sum m_\nu)$. Right plot: 2D confidence level contours for  $(\sigma_8, \sum m_\nu)$.
}
\label{fig:CosmoMnuLyaPlanck}
\end{center}
\end{figure}

Fig.~\ref{fig:CosmoMnuLyaPlanck}  shows that the combination of Planck and Ly$\alpha$ is a very efficient way of constraining cosmological parameters, especially $\sum m_\nu$.  As we explained in PY15,  Ly$\alpha$ data constrain $\Omega_m$ and $\sigma_8$ largely independently of $\sum m_\nu$ because they have different
impacts on the shape of the power spectrum. On the other hand, in CMB data,  $\sum m_\nu$ is strongly correlated with $\Omega_m$ and $\sigma_8$. For the Planck constraints, high $\sum m_\nu$ corresponds to low $\sigma_8$ because of the suppression of power on
small scales by neutrino free streaming.  The positive correlation between $\Omega_m$ and $\sum m_\nu$ is more subtle: with $\Omega_c h^2$ and $\Omega_b h^2$ well constrained by the acoustic peaks, raising $\sum m_\nu$ increases the matter density at low redshift after neutrinos become non-relativistic,
and within $\Lambda$CDM this requires a decrease in $h$ to maintain the well determined angular diameter distance to last scattering,
and this in turn corresponds to higher $\Omega_m$(see, e.g., \S 6.4 of \cite{Planck2015}). The end result is that the \lya\ and Planck contours intersect only near $\sum m_\nu=0$. By combining  (BOSS and XQ-100) Ly$\alpha$ and Planck TT+lowP data, we  constrain $\sum m_\nu$ to be less than 0.14~eV at 95\% C.L.. Finally, adding polarization or BAO to the Planck + Ly$\alpha$ contours does not lead to significant further improvement of the constraints  as shown in column (5) of Tab~\ref{tab:fit_standard_nu}.

\begin{figure}[htbp]
\begin{center}
\epsfig{figure= 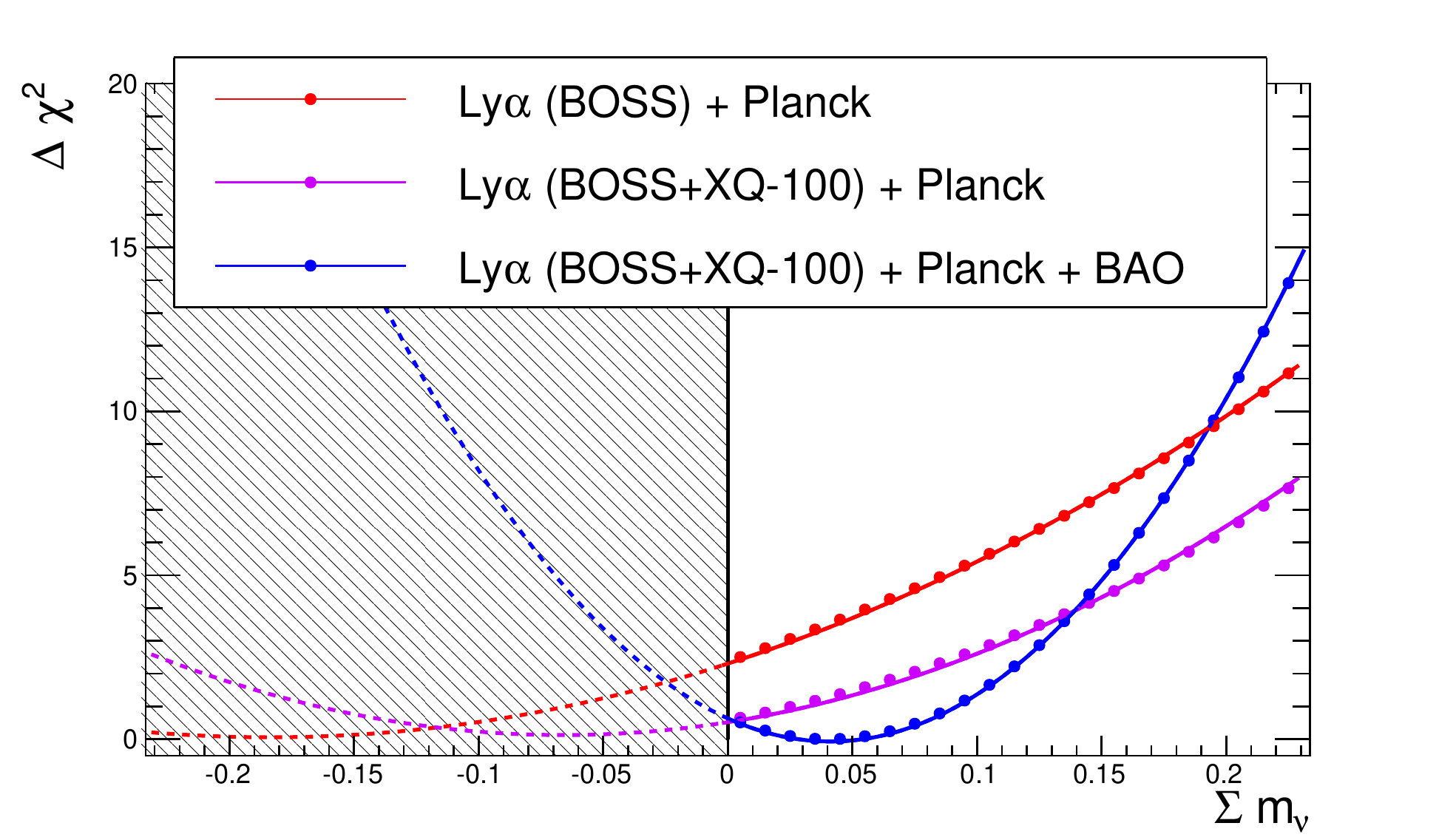,width = 12cm} 
\caption{\it  $\Delta \chi^2$ profile as a function of $\sum m_\nu$ for the three configurations (Ly$\alpha$ (BOSS) + Planck),  (Ly$\alpha$ (BOSS+XQ-100) + Planck) and (Ly$\alpha$ (BOSS+XQ-100) + Planck + BAO). Each point is the $\Delta \chi^2$ obtained after a maximization of the total likelihood over the other free parameters.  The points are fitted by a parabola and extrapolated into the negative region as proposed in~\cite{PlanckCollaboration2014Freq}.
}
\label{fig:Chi2ScanMnu}
\end{center}
\end{figure}

The current limit on $\sum m_\nu$ is a bit looser that the 0.12~eV bound of PY15, whereas we note a small improvement in the Ly-$\alpha$ alone constraint with the inclusion of XQ-100  (cf. Sec.~\ref{sec:LyaAlone}). The origin of this result can be understood when plotting the $\chi^2$ profile for three combinations of data sets  (cf. Fig.~\ref{fig:Chi2ScanMnu} (right)). The minimum for  the BOSS Ly$\alpha$ + Planck configuration (red curve) occurs for $\sum m_\nu<0$. The fact that the CMB data sets have their minimum in the unphysical (negative  $\sum m_\nu$) region was already  discussed in~\cite{PlanckCollaboration2014Freq}. In the results presented in Tab.~\ref{tab:fit_standard_nu},  the limit on the total neutrino mass is derived by computing the probability of $\Delta \chi^2 (\sum m_\nu) =\chi^2(\sum m_\nu) -\chi^2(\sum m_\nu=0)$ with one degree of freedom.  In the case of the (BOSS+XQ-100) Ly$\alpha$ + Planck configuration, the curvature of the $\chi^2$ profiles stays almost unchanged, but the position of the minimum is shifted closer to   the physical ($\sum m_\nu>0$) region, causing a looser bound on  $\sum m_\nu$.

The shift of  the $\chi^2$  minimum into the physical region can be easily explained. By adding XQ-100, the combined value of $\sigma_8$ is smaller compared to BOSS alone (see first and second columns of Tab~\ref{tab:fit_standard_nu}). The two parameters $\sigma_8$ and $\sum m_\nu$ are anti-correlated for Planck data as shown on Fig.~\ref{fig:Chi2ScanMnu} (left). As a consequence, XQ-100 pushes toward higher $\sum m_\nu$.

 Finally, as shown in   Fig.~\ref{fig:Chi2ScanMnu} (right) (blue curve),  adding  BAO data  definitively forces the minimum  back into the physical ($\sum m_\nu>0$) region. Hence, despite a much more constraining set of data (the limit at 3$\,\sigma$ or more is the tightest when all Ly$\alpha$, CMB and BAO data are included), we obtain a limit on the total neutrino mass of $\sum m_\nu < 0.14$~eV at 95\% CL, identical to the limit with Ly$\alpha$ and Planck.


\section{{Constraints on  \texorpdfstring{$\Lambda$WDM}{WDM}}}
\label{sec:WDM}
In this section, we present the cosmological results of our analysis on $\Lambda$WDM cosmology with thermal relics and non-resonantly produced neutrinos. We first extend the method explained in BP16 to  XQ-100 data. We then compare the constraints obtained with the Ly$\alpha$ power spectrum measured in this study to those obtained with the power spectrum of~\cite{Irsic2016}. 

\subsection{\texorpdfstring{$\Lambda$WDM}{WDM} cosmology  with XQ-100}
\label{sec:WDMmyXQ100}
To probe $\Lambda$WDM cosmology, we follow a similar  approach  as described in Sec.\ref{sec:method}. The only significant difference is related to the uncertainty on the reionization history of the universe.  As the redshift at which the UV background onsets affects the Jeans smoothing scale of the baryon gas~\cite{Gnedin&Hui98} in a manner similar to the free streaming scale of warm dark matter particles,  altering the reionization redshift $z_{\star}$  impacts the definition of the WDM free streaming scale. Fig.~13 of~\cite{McDonald2005} shows that an increase in the redshift of reionization from $z_{\star}=7$ to 17 suppresses the Ly$\alpha$ flux power spectrum in the largest  $k$-modes present in the BOSS data ($k \sim 0.02\, \rm{s\,km^{-1}}$) by about 1\% at $z=2.1$ and 4\% at $z=4.0$.  For  XQ-100 where the power spectrum is measured at much smaller scales, ($k \sim 0.07 \, \rm{s\,km^{-1}}$ at $z=4.0$), we are very sensitive to such an effect.  Using the study of reionization by~\cite{McDonald2005}, we model the effect of reionization over the power spectrum and we introduce a nuisance parameter representing $z_{\star}$. This new parameter is let free in the likelihood with a constraint $z_{\star} = 9.0 \pm 1.5$.   The central value and range of this external constraint are defined in order to encompass the most recent measurements of the redshift of reionization~\cite{Hinshaw2013,Planck2015,Planck2016PolarReio}.

\begin{figure}[htbp]
\begin{center}
\epsfig{figure= 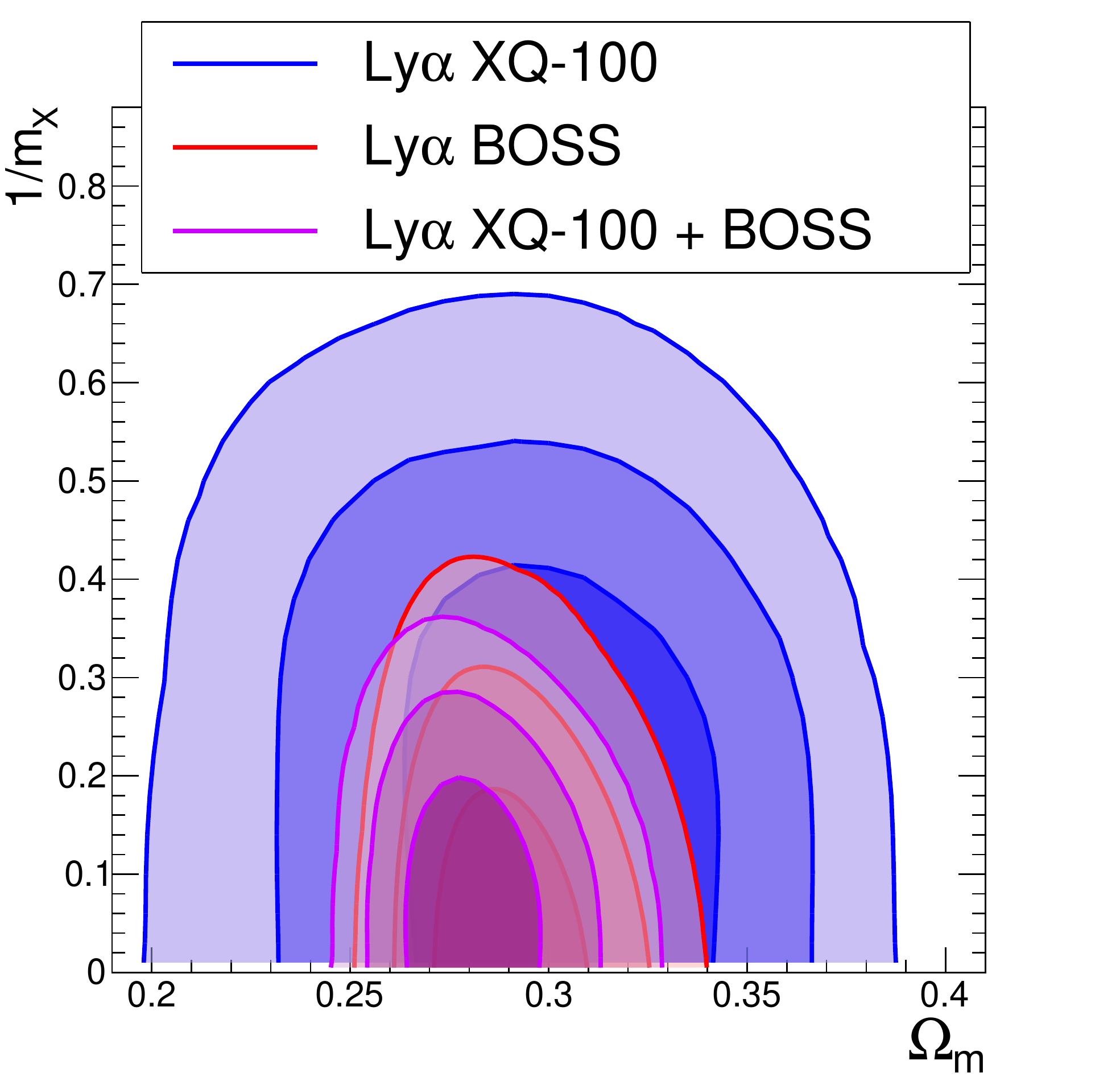,width = 7.5cm}
\epsfig{figure= 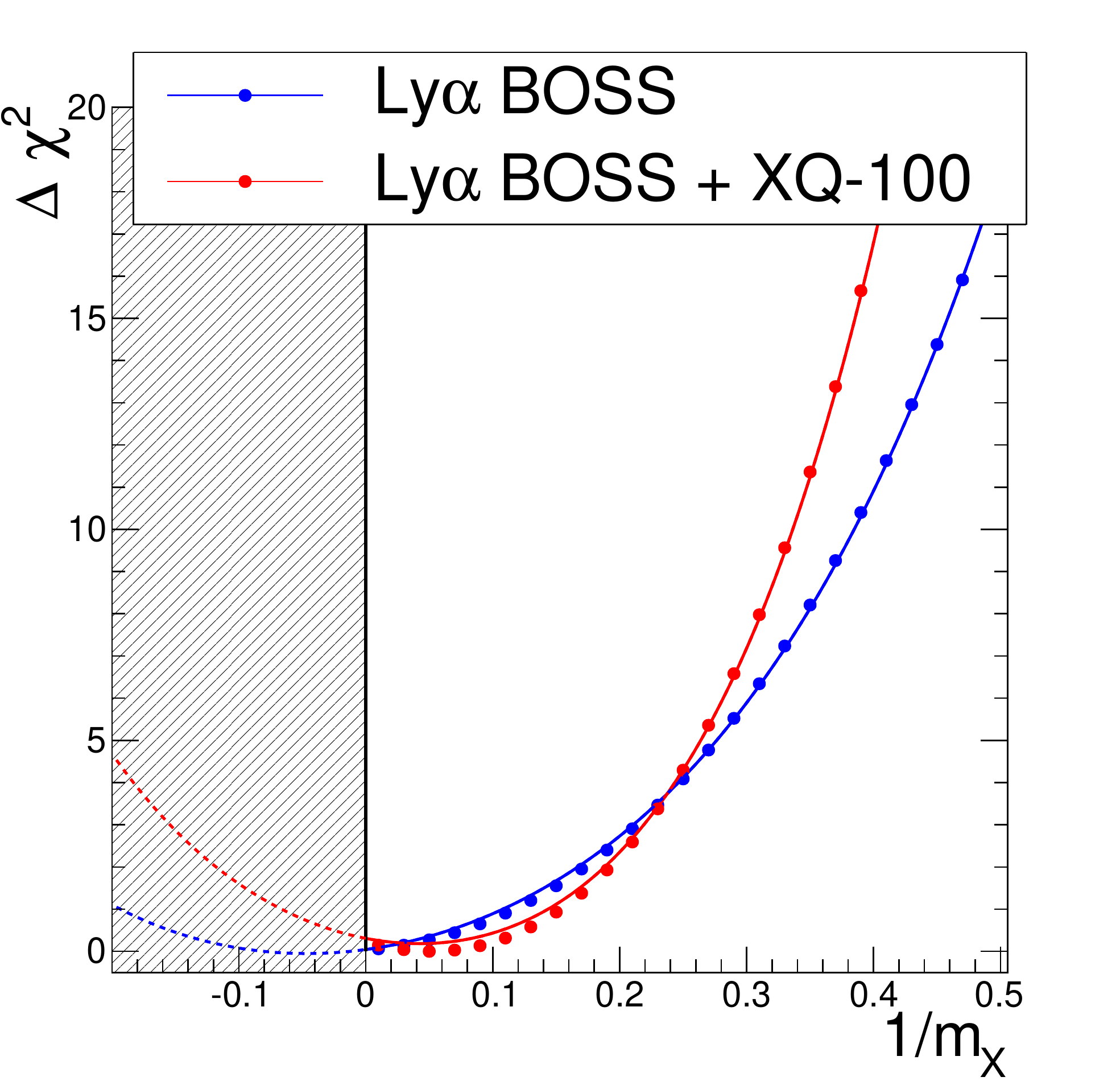,width = 7.5cm}
\caption{\it  Left plot : 2D confidence level contours for  the $(\Omega_m , 1/m_{x})$  cosmological parameters.  The 68\%, 95\% and 99.7\% confidence contours are obtained  for the XQ-100 Ly$\alpha$ data, for the BOSS Ly$\alpha$ data separately and for the combination of XQ-100 and BOSS.  Right plot: $\Delta \chi^2$ profile as a function of $1/m_x$ for the two configurations: BOSS alone and the combination of BOSS and XQ-100. Each point is the $\Delta \chi^2$ obtained after a maximization of the total likelihood over the other free parameters.}
\label{fig:OmegamMx}
\end{center}
\end{figure}

A fit to the power spectrum of XQ-100 Ly$\alpha$ forest, assuming the expansion rate value  $H_0 = 67.3 \pm 1.0 \; \rm{km~s^{-1}~Mpc^{-1}}$ issued by~\cite{Planck2015}, yields a lower bound  $m_X > 2.08 \, \rm{keV}$ (95\% C.L.) for thermal relics and $m_s > 10.2 \: \rm{keV}$ for Dodelson-Widrow~\cite{DodelsonWidrow94} sterile neutrinos (95\% CL). These bounds are roughly twice lower than the bounds given with BOSS alone by BP15 in Fig.~\ref{fig:OmegamMx} (left) and Tab.~\ref{tab:Limit_WDM}.

When combining XQ-100 with BOSS Ly$\alpha$ forest power spectra, the 95\% C.L. limit is slightly improved compared to the one from BOSS alone. It increases from $m_X \gtrsim 4.09 \: \rm{keV}$ to $m_X \gtrsim 4.17 \: \rm{keV}$. The  $3~\sigma$ bound shows a more significantly improvement, increasing from $m_X \gtrsim 2.74 \: \rm{keV}$ for BOSS alone to $m_X \gtrsim 3.10 \: \rm{keV}$ for the combined BOSS+XQ-100 data set.

\begin{figure}[htbp]
\begin{center}
\epsfig{figure= 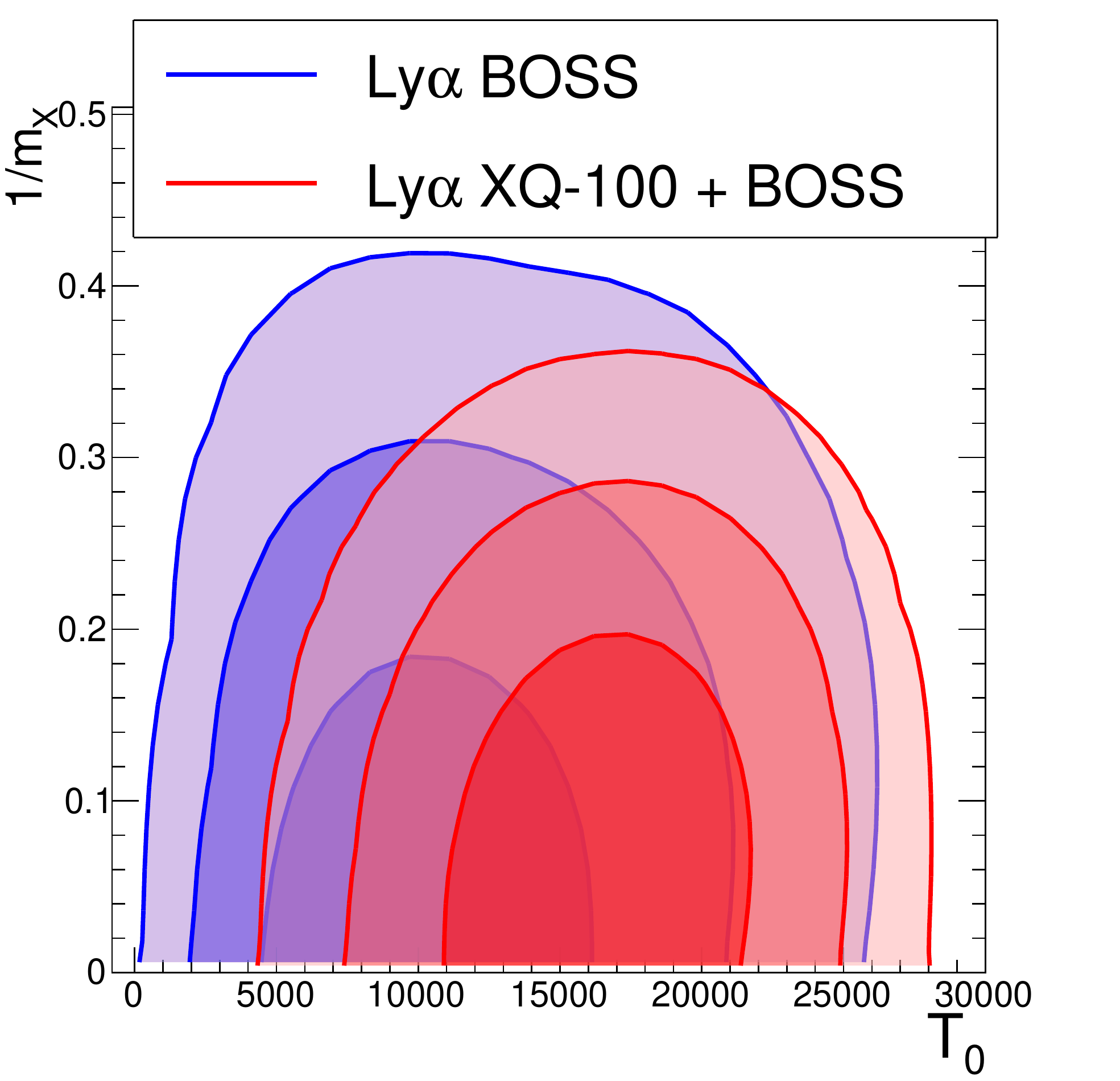,width = 7.5cm}
\epsfig{figure= 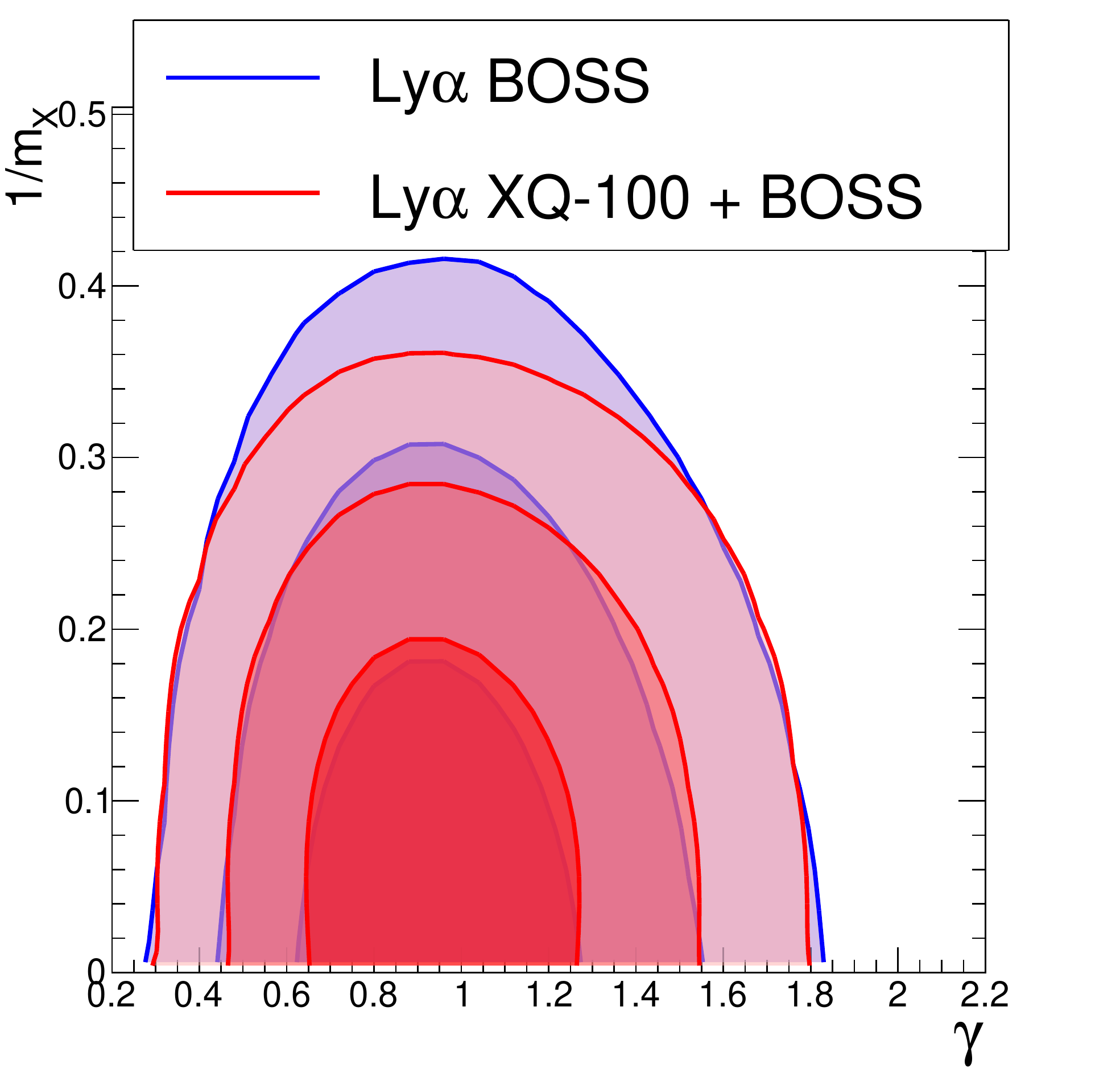,width = 7.5cm}
\caption{\it 2D confidence level contours between IGM temperature parameters and $1/m_x$. The 68\%, 95\% and 99.7\% confidence contours are obtained  for the BOSS Ly$\alpha$ data alone and for the combination of XQ-100 and BOSS.  Left plot :  $(T_0 , 1/m_{x})$.  Right plot:    $(\gamma , 1/m_{x})$.   }
\label{fig:T0Mx}
\end{center}
\end{figure}

The reason of this improvement is clearly illustrated in Fig.~\ref{fig:OmegamMx} (right), which shows the $\chi^2$ profile for two combinations of data sets: BOSS alone and BOSS+XQ-100. Clearly, the curvature increases by adding XQ-100  but the position of the minimum is shifted into   the physical ($1/m_x>0$) region, explaining a small improvement of the 95\% C.L bound and a larger improvement at higher significance ($3~\sigma$ or more). Fitting the $\chi^2$ profile by $\Delta \chi^2 (1/m_X) =\chi^2_0 + (1/m_x-1/m_{x_0})^2/\sigma^2+\alpha\cdot(1/m_x-1/m_{x_0})^4$, the parameter $\sigma$ provides an estimator of the statistical sensitivity on $1/m_X$. The combination with XQ-100 allows us to reduce $\sigma$ from 0.15 to 0.12, representing a  25\% gain in statistical sensitivity.

Although we believe that a Gaussian constraint with a sigma of 1.5 on $z_{\star}$ allows us to encompass the range of allowed $z_{\star}$ from CMB results (in particular~\cite{Hinshaw2013,Planck2015,Planck2016PolarReio}), we  released the constraint on  $z_{\star}$, allowing for a wider variation range, to study the impact on the warm dark matter mass bound. The effect is small: If we increase the sigma to 2.5, the 95\% CL limit on $m_X$ decreases from $4.17 \: \rm{keV}$ to  $3.90 \: \rm{keV}$.

Finally, it has been recently argued in \cite{Garzilli2015} that the small-scale cutoff in the power spectrum can be accounted for by a warm IGM rather than a warm DM particle. By adding XQ-100 to BOSS data, in Fig~\ref{fig:T0Mx}, we do not observe any significant change  in the IGM temperature $T_0$ nor in  the logarithmic slope $\gamma$ of the dependence of the IGM temperature with overdensity $\delta$.  Additional Ly$\alpha$ forest data at higher redshifts ($z\ge 4.5$) are needed to better study  this hypothesis.

In summary, with the inclusion of XQ-100 in addition to BOSS  into the analysis of BP16, we confirm our original limit from BOSS alone, in the case of $\Lambda$WDM model. We issue tighter bounds on pure dark matter particles: $m_X \gtrsim 4.17 \: \rm{keV}$ (95\% C.L.) for early decoupled thermal relics  and its corresponding bound for a non-resonantly produced right-handed neutrino $m_s \gtrsim 25.0 \: \rm{keV}$ (95\% C.L.).

\subsection{Comparison with Irsic et al. \texorpdfstring{$\rm{Ly}\alpha$}{Lya}  power spectrum}

\begin{figure}[htbp]
\begin{center}
\epsfig{figure= 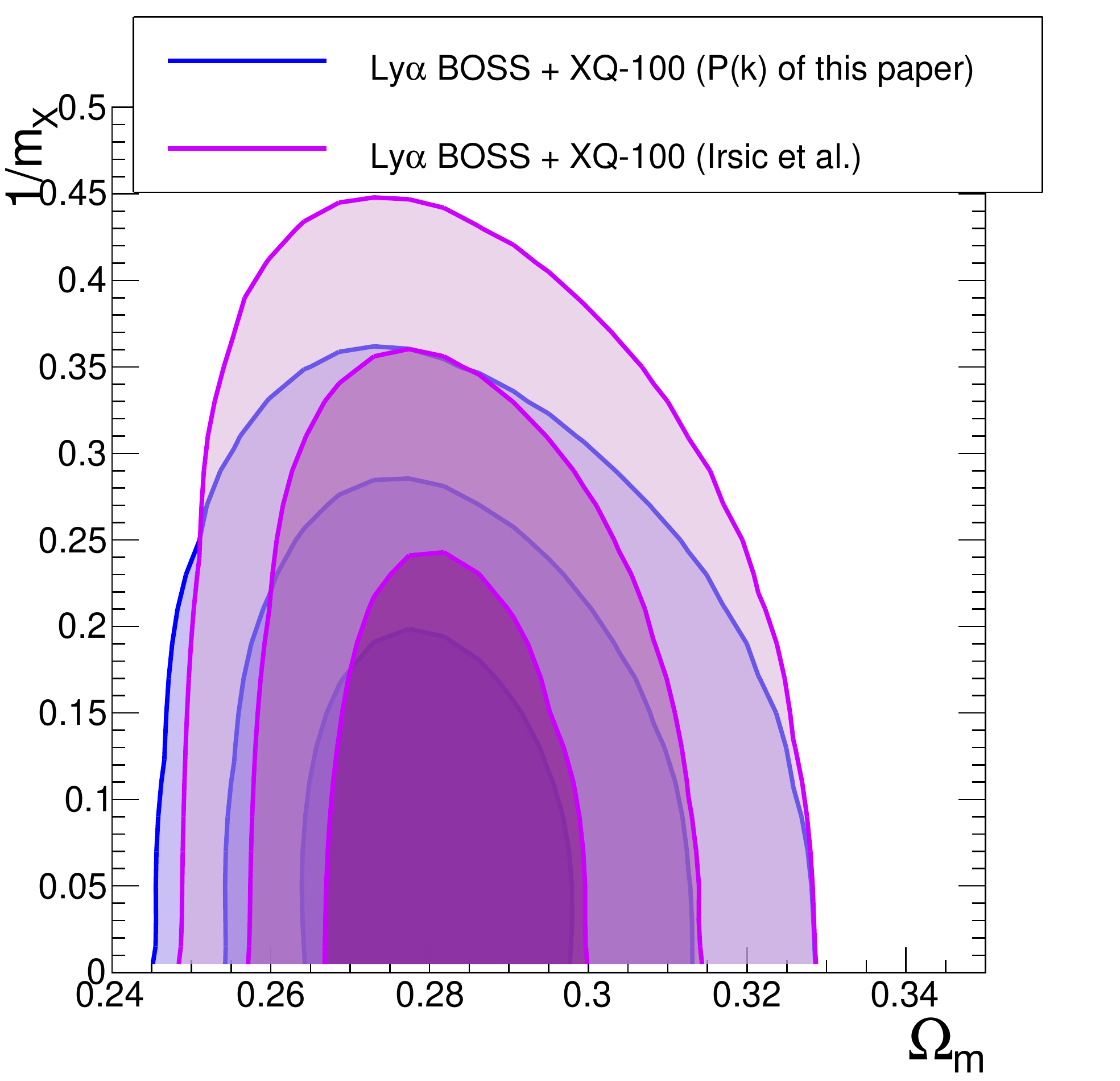,width = 7.5cm}
\epsfig{figure= 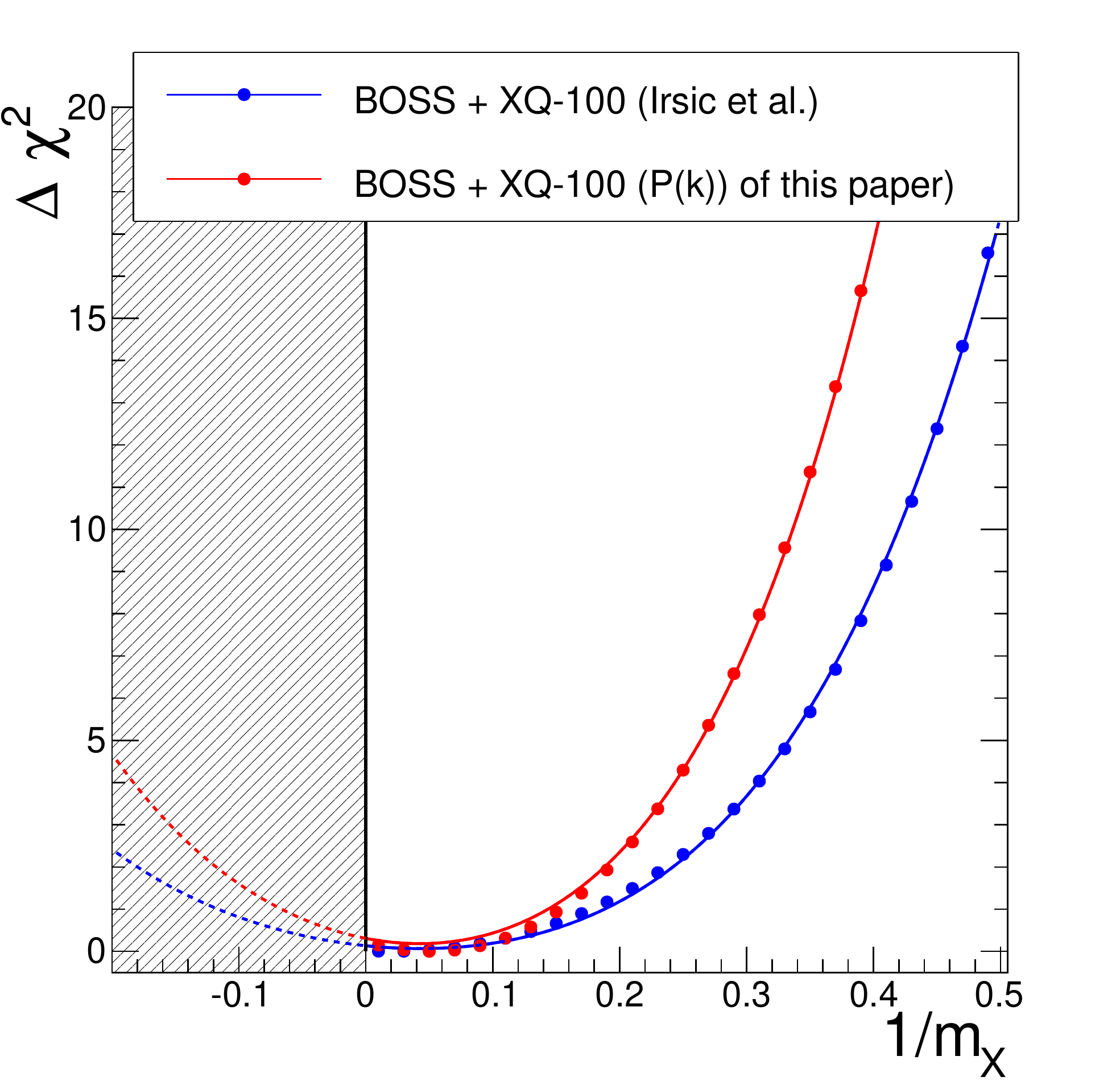,width = 7.5cm}
\caption{\it  Comparison of the results obtained with the $P(k)$ computed in this paper and the $P(k)$ measured in~\cite{Irsic2016}. Left plot : 2D confidence level contours in  the $(\Omega_m , 1/m_{x})$  plane.  The 68\%, 95\% and 99.7\% confidence contours are obtained  for the combination of XQ-100 and BOSS data. Right plot: $\Delta \chi^2$ profile as a function of $\sum m_\nu$ for the combination of BOSS and XQ-100. Each point is the $\Delta \chi^2$ obtained after a maximization of the total likelihood over the other free parameters. }
\label{fig:OmegamMx_Irsic}
\end{center}
\end{figure}

Recently, the XQ-100 team  released the measurement of the Ly$\alpha$ forest power spectrum  in~\cite{Irsic2016}. Applying the methodology of BP16  to their XQ-100 power spectrum alone, we find a bound at  $m_X > 1.90 \: \rm{keV}$ (95\% C.L.) for thermal relics, which is fully compatible with our bound, $m_X > 2.08 \: \rm{keV}$ (95\% C.L.).  We suspect that the slightly tighter limit we obtain comes from the better estimation of the resolution as explained in Sec.~\ref{sec:reso}.   By removing the highest $k$-bins, we indeed observe a small trend that may explain partially the difference. For instance, when we do not take into account the $k$-bins between 0.06 and $0.07\:{\rm (km/s)}^{-1}$ for the highest redshift bin, the limit decreases from 2.08 to $2.03\: \rm{keV}$. Tab.~\ref{tab:Limit_WDM} and Fig.~\ref{fig:OmegamMx_Irsic} summarize the comparison on  the 95\% confidence limits obtained with the two power spectra. 

\begin{table}[htbp]
\caption{\it 95\% confidence limit on $m_X$, the mass of thermal relics, and $m_s$, the mass of  non-resonantly produced neutrinos, in a WDM scenario. The first row corresponds to the most stringent lower limit on WDM mass to date prior to this work, obtained with BOSS data in BY16. The other lines are for this work.}
\begin{center}
\begin{tabular}{lcc}
\hline\hline
& $P(k)$ of this paper & $P(k)$ of~\cite{Irsic2016} \\
\hline
BOSS alone~\cite{Baur2016}  & \multicolumn{2}{c}{    $m_X \gtrsim 4.09 \: \rm{keV}$   and       $m_s \gtrsim 24.4 \: \rm{keV}$}  \\
\hline
XQ-100 alone   &  $m_X \gtrsim 2.08 \: \rm{keV}$  $m_s \gtrsim 10.2 \: \rm{keV}$ &    $m_X \gtrsim 1.90 \: \rm{keV}$  $m_s \gtrsim 9.0 \: \rm{keV}$  \\
\hline
XQ-100 + BOSS & $m_X \gtrsim 4.17 \: \rm{keV}$  $m_s \gtrsim 25.0 \: \rm{keV}$ &    $m_X \gtrsim 3.29 \: \rm{keV}$  $m_s \gtrsim 18.4 \: \rm{keV}$  \\
\hline
\end{tabular}
\end{center}
\label{tab:Limit_WDM}
\end{table}

\subsection{Adding HIRES/MIKE power spectrum}

The analysis presented in~\cite{Irsic2017} shows that the combination of the XQ-100 and HIRES/MIKE datasets can significantly improve the limit on $m_X$.  Indeed, the two datasets have different degeneracies between astrophysical and cosmological parameters that are disentangled when the data are combined, thanks to the higher resolution of the HIRES/MIKE spectrograph. By studying  the three datasets BOSS, XQ-100 and HIRES/MIKE, we can thus expect an improvement over BOSS+XQ-100. However, as noted before, since the  first snapshot of our simulations is taken at $z=4.6$, we can only use the lowest two redshift bins of HIRES/MIKE   ($z=4.2$ and $z=4.6$) .

\begin{figure}[htbp]
\begin{center}
\epsfig{figure= 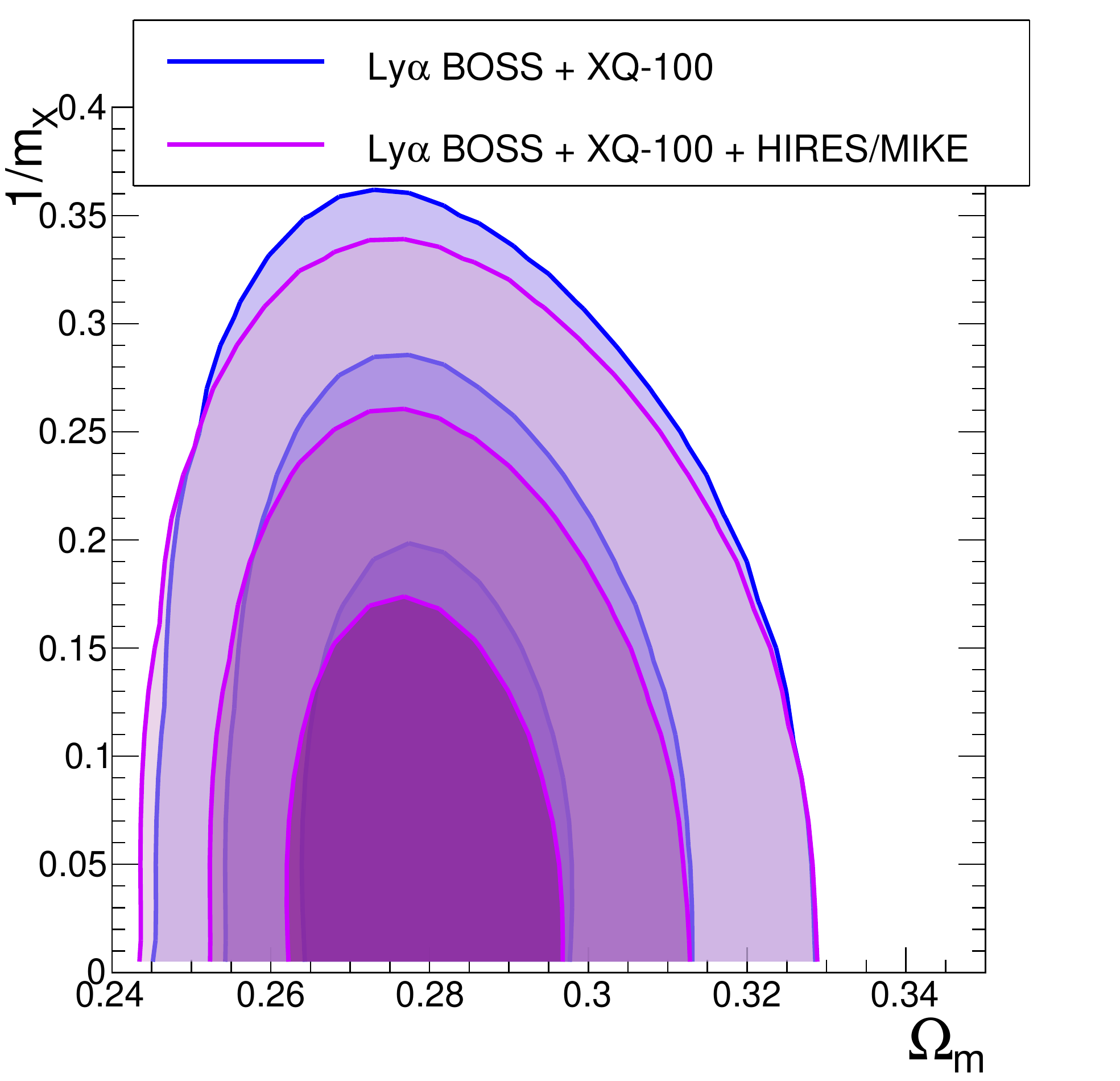,width = 7.5cm}
\epsfig{figure= 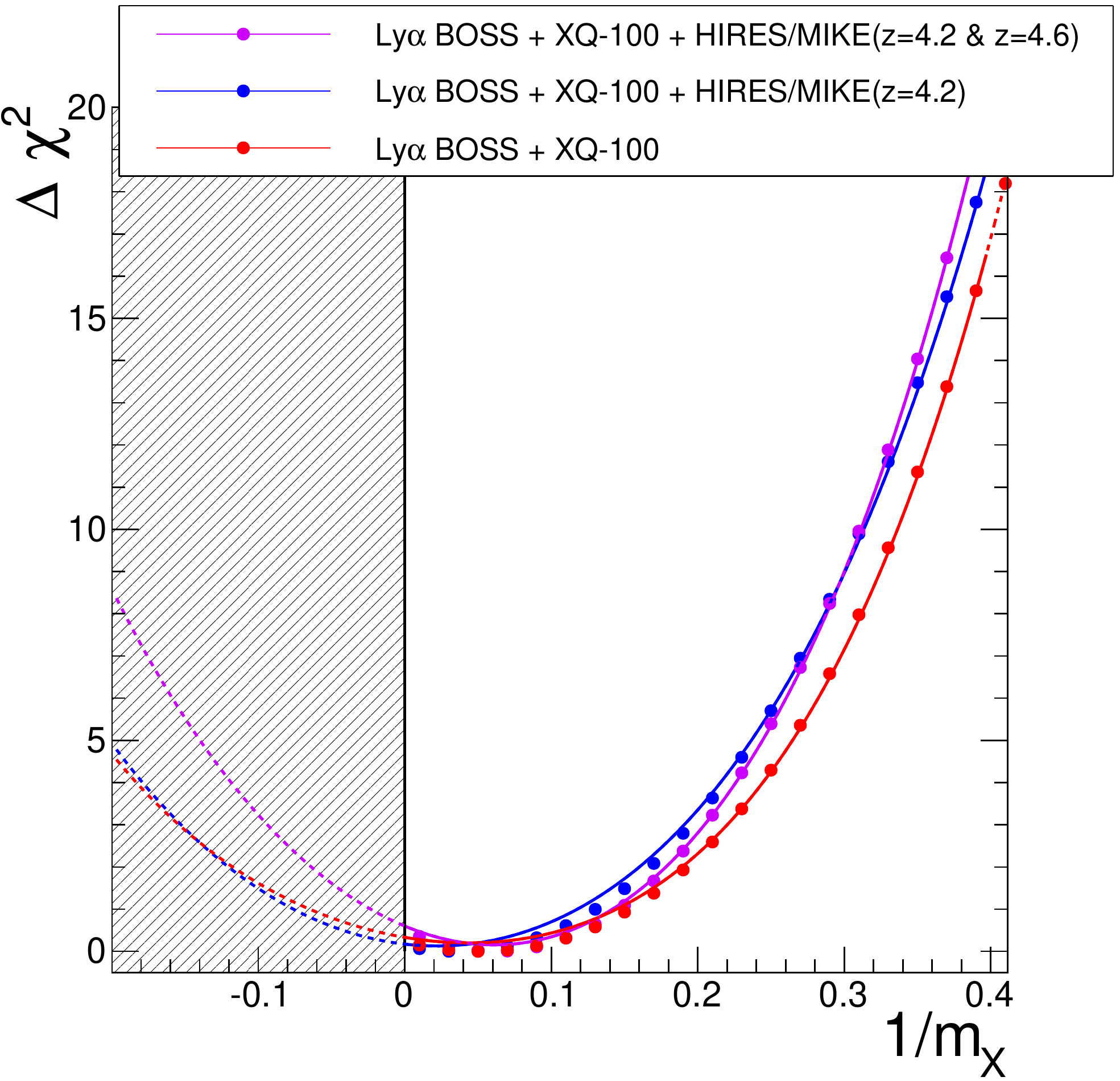,width = 7.5cm}
\caption{\it  Left plot : 2D confidence level contours for  the $(\Omega_m , 1/m_{x})$  cosmological parameters.  The 68\%, 95\% and 99.7\% confidence contours are obtained  for the combination of XQ-100 and BOSS and with adding HIRES/MIKE power spectrum. Right plot: $\Delta \chi^2$ profile as a function of $1/m_x$ for the three configurations: the combination of (BOSS + XQ-100), (BOSS + XQ-100 + HIRES/MIKE with one redshift bin $z=4.2$ ) and (BOSS + XQ-100 + HIRES/MIKE with two redshift bins($z=4.2$ and $z=4.6$)). Each point is the $\Delta \chi^2$ obtained after a maximization of the total likelihood over the other free parameters.}
\label{fig:OmegamMx_HIRES}
\end{center}
\end{figure}

Fig.~\ref{fig:OmegamMx_HIRES} (left) illustrates the  improvement in the $1/m_X - \Omega_m$ plane. The new bounds with HIRES/MIKE are $m_X \gtrsim 4.65 \: \rm{keV}$ and $m_s \gtrsim 28.8  \: \rm{keV}$ (95\% C.L.).  As we explained in Sec.~\ref{sec:WDMmyXQ100}, the fit of  the $\chi^2$ profile shown on Fig.~\ref{fig:OmegamMx_HIRES} (right) by a quadratic term demonstrates a reduction of $\sigma$ from 0.123 to 0.105 and finally 0.093 when we add successively the $z=4.2$  and $z=4.6$ redshift bins of HIRES/MIKE. It represents a  gain of respectively 17\% and 13\%  in statistical sensitivity. In total,  it is a gain in statistical sensitivity of 60\% with respect to BOSS alone. We do not observe such an impressive gain for the 95\% CL bound because the $\chi^2$ minimum moves from the negative (unphysical) region to the physical region. Our final limit on $m_X$ is in perfect agreement with the new bound $m_X \gtrsim 5.3 \: \rm{keV}$  given in~\cite{Irsic2017}, confirming the consistent bounds measured by the various teams~\cite{SMT08,Viel2013,Baur2016,Irsic2017}.


\section{Conclusion}
\label{sec:conclusion}

We measure the 1D Ly$\alpha$ power spectrum of the 100 quasars of the XQ-100 survey,     applying the method~\cite{Palanque-Delabrouille2013}  developed  for BOSS. The power spectrum is computed with a Fourier transform. We  use the flux error measured by the official pipeline~\cite{Lopez2016} to estimate the noise power spectrum. Since we aim at measuring  small scales that are crucial to constrain WDM, the determination of the spectrograph resolution was subject to a special dedicated treatment. Using the raw 2D quasar spectra, we  checked  that for many observations, the seeing is better than the slit width. Therefore, our computation of the spectrograph resolution relies upon the measured seeing at VLT instead of upon the slit width. We  measured the power spectrum for three redshift bins  covering the \ion{H}{i} absorption range  $3.0<z<4.2$ and reaching $k$-modes  as high as $0.070\, {\rm  s\,km^{-1} }$.

Fitting Ly$\alpha$ data alone leads to cosmological parameters in excellent agreement with the values derived independently from CMB data. Combining BOSS Ly$\alpha$ and XQ-100 Ly$\alpha$, we  constrain the sum of neutrino masses to $\sum m_\nu < 0.8$~eV (95\% C.L.) including all identified systematic uncertainties. With the addition of CMB data, this bound  is tightened to $\sum m_\nu < 0.14$~eV (95\% C.L.).

With their sensitivity to small scales, Ly$\alpha$ data are ideal to constrain $\Lambda$WDM models. Using 
 XQ-100 alone, we issue lower bounds on pure dark matter particles: $m_X \gtrsim 2.08 \: \rm{keV}$ (95\% C.L.) for early decoupled thermal relics, and $m_s \gtrsim 10.2 \: \rm{keV}$ (95\% C.L.)  for non-resonantly produced right-handed neutrinos.  Combining the 1D  Ly$\alpha$-forest power spectrum measured by BOSS and XQ-100, we improve the two bounds to $m_X \gtrsim 4.17 \: \rm{keV}$ and $m_s \gtrsim 25.0 \: \rm{keV}$ (95\% C.L.), slightly more constraining  than what was  achieved in Baur et al. 2015~\cite{Baur2016}  with BOSS data alone. The  $3~\sigma$ bound shows a  more significant improvement, increasing from $m_X \gtrsim 2.74 \: \rm{keV}$ for BOSS alone to $m_X \gtrsim 3.10 \: \rm{keV}$ for the combined BOSS+XQ-100 data set. Finally, the addition of the higher-resolution HIRES/MIKE power spectrum at redshifts $z=4.2$ and 4.6 allows us to  further improve the two limits to $m_X \gtrsim 4.65 \: \rm{keV}$ and $m_s \gtrsim 28.8  \: \rm{keV}$ (95\% C.L.).

\acknowledgments

We acknowledge PRACE (Partnership for Advanced Computing in Europe) for awarding us access to resources Curie thin nodes and Curie fat nodes, based in France at TGCC, under allocation numbers 2010PA2777, 2014102371 and 2012071264. 
This work was also granted access to the resources of CCRT under the allocation 2013-t2013047004 made by
GENCI (Grand Equipement National de Calcul Intensif).
The authors thank Valentina D'Odorico, Patrick Petitjean and Pasquier Noterdaeme for useful discussions on XSHOOTER. The authors are also grateful to Vid Irsic and Matteo Viel for discussions on XQ-100 Ly$\alpha$ power spectrum and for providing HIRS/MIKE power spectrum. Finally, the authors wish to thank the organizers of the {\it Neutral H for cosmology} workshop held in Berkeley in January 2017, which triggered this work.\\

\newpage
\bibliographystyle{unsrtnat_arxiv}
\bibliography{biblio}

\section*{Appendix: Measured XQ-100  \texorpdfstring{$\rm{Ly}\alpha$}{Lya}  flux power spectrum}
\label{sec:appendix}
Table~\ref{tab:Pk} gives the values of the power spectra from XQ-100 quasar data presented in this work.  The method used to estimate  the spectrograph resolution from the seeing of the observations, and the noise power spectrum from the pipeline pixel uncertainties,   are described in section~\ref{sec:reso}. The columns list the redshift, the $k$ mode, the flux power spectrum $P_{1D}(k)$, the statistical uncertainty $\sigma_{\rm stat}$ and the noise power spectrum $P^{noise}(k)$.  The 1D power spectrum, $P_{1D}(k)$ is computed as explain in Eq.~\ref{eq:P1D_FFT}, in particular the uncorrelated component related to metal absorption is subtracted.  

\begin{table}[h]
\caption{1D Flux power spectrum from the XQ-100 survey, for the three redshift bins of this work.  }
\label{tab:Pk}
\end{table}

\begin{table}[p]
  \scriptsize
  \begin{center}
  \begin{tabular}{ccccc}
  \hline
    \hline
    $z$ & $k\: {\rm [s\,km^{-1}]}$ &  $P_{1D}(z,k)\: {\rm [km\,s^{-1}]}$ &  $\sigma_{\rm stat}(z,k)\: {\rm [km\,s^{-1}]}$ & $P^{noise}(z,k)\: {\rm [km\,s^{-1}]}$ \\
\hline
 3.200000 & 0.001500 & 58.113500 & 4.643160 & 0.025454 \\ 
3.200000 & 0.002500 & 54.555800 & 3.931500 & 0.024762 \\ 
3.200000 & 0.003500 & 44.943600 & 3.079560 & 0.023012 \\ 
3.200000 & 0.004500 & 42.804400 & 3.401900 & 0.024050 \\ 
3.200000 & 0.005500 & 38.448600 & 3.022150 & 0.022354 \\ 
3.200000 & 0.006500 & 32.178500 & 2.667050 & 0.024910 \\ 
3.200000 & 0.007500 & 30.951600 & 2.094940 & 0.023047 \\ 
3.200000 & 0.008500 & 30.244900 & 2.215940 & 0.025844 \\ 
3.200000 & 0.009500 & 28.993000 & 2.180230 & 0.024560 \\ 
3.200000 & 0.010500 & 25.604600 & 1.596620 & 0.024416 \\ 
3.200000 & 0.011500 & 22.290900 & 1.532780 & 0.025576 \\ 
3.200000 & 0.012500 & 24.299100 & 1.857580 & 0.024284 \\ 
3.200000 & 0.013500 & 19.278000 & 1.491770 & 0.025975 \\ 
3.200000 & 0.014500 & 18.835600 & 1.197180 & 0.027148 \\ 
3.200000 & 0.015500 & 16.788400 & 1.272570 & 0.027732 \\ 
3.200000 & 0.016500 & 19.063900 & 1.502010 & 0.026140 \\ 
3.200000 & 0.017500 & 17.625900 & 1.351370 & 0.026887 \\ 
3.200000 & 0.018500 & 14.538700 & 0.993207 & 0.027246 \\ 
3.200000 & 0.019500 & 16.092300 & 1.105720 & 0.028970 \\ 
3.200000 & 0.020500 & 14.805700 & 1.084320 & 0.027306 \\ 
3.200000 & 0.021500 & 15.687800 & 1.061050 & 0.026487 \\ 
3.200000 & 0.022500 & 12.796500 & 0.795661 & 0.028633 \\ 
3.200000 & 0.023500 & 15.005400 & 1.464770 & 0.030546 \\ 
3.200000 & 0.024500 & 13.566500 & 1.060040 & 0.030663 \\ 
3.200000 & 0.025500 & 13.100100 & 1.069790 & 0.030242 \\ 
3.200000 & 0.026500 & 10.965000 & 0.839604 & 0.031844 \\ 
3.200000 & 0.027500 & 9.597180 & 0.781718 & 0.032698 \\ 
3.200000 & 0.028500 & 10.595200 & 0.961049 & 0.032106 \\ 
3.200000 & 0.029500 & 9.824680 & 0.723203 & 0.034216 \\ 
3.200000 & 0.030500 & 9.827700 & 0.833807 & 0.037026 \\ 
3.200000 & 0.031500 & 9.041100 & 0.688367 & 0.035163 \\ 
3.200000 & 0.032500 & 8.845860 & 0.723757 & 0.036374 \\ 
3.200000 & 0.033500 & 8.703450 & 0.767202 & 0.036836 \\ 
3.200000 & 0.034500 & 8.763500 & 0.695326 & 0.041306 \\ 
3.200000 & 0.035500 & 7.480200 & 0.660709 & 0.041548 \\ 
3.200000 & 0.036500 & 8.022700 & 0.640697 & 0.038998 \\ 
3.200000 & 0.037500 & 7.967960 & 0.760975 & 0.043356 \\ 
3.200000 & 0.038500 & 7.814740 & 0.541527 & 0.040360 \\ 
3.200000 & 0.039500 & 7.582290 & 0.656729 & 0.043170 \\ 
3.200000 & 0.040500 & 6.143250 & 0.401928 & 0.044032 \\ 
3.200000 & 0.041500 & 5.263360 & 0.366231 & 0.048178 \\ 
3.200000 & 0.042500 & 6.010660 & 0.528384 & 0.051034 \\ 
3.200000 & 0.043500 & 5.254620 & 0.340819 & 0.047227 \\ 
3.200000 & 0.044500 & 5.063790 & 0.392426 & 0.051594 \\ 
3.200000 & 0.045500 & 5.343530 & 0.408905 & 0.057472 \\ 
3.200000 & 0.046500 & 5.146310 & 0.405373 & 0.055439 \\ 
3.200000 & 0.047500 & 4.945440 & 0.353022 & 0.060523 \\ 
3.200000 & 0.048500 & 4.670380 & 0.342687 & 0.063460 \\ 
3.200000 & 0.049500 & 5.446050 & 0.576584 & 0.061197 \\ 
3.200000 & 0.050500 & 4.889070 & 0.436336 & 0.068134 \\ 
3.200000 & 0.051500 & 3.878910 & 0.304877 & 0.068175 \\ 
3.200000 & 0.052500 & 3.709630 & 0.333078 & 0.074124 \\ 
3.200000 & 0.053500 & 3.853920 & 0.358999 & 0.080345 \\ 
3.200000 & 0.054500 & 3.654200 & 0.277982 & 0.074647 \\ 
3.200000 & 0.055500 & 3.731370 & 0.342549 & 0.084883 \\ 
3.200000 & 0.056500 & 3.197610 & 0.267283 & 0.091170 \\ 
3.200000 & 0.057500 & 3.677780 & 0.340790 & 0.097883 \\ 
3.200000 & 0.058500 & 2.982880 & 0.233127 & 0.094032 \\ 
3.200000 & 0.059500 & 2.565260 & 0.206981 & 0.104943 \\ 
3.200000 & 0.060500 & 2.700700 & 0.232480 & 0.113155 \\ 
3.200000 & 0.061500 & 3.348200 & 0.295818 & 0.111648 \\ 
3.200000 & 0.062500 & 2.276380 & 0.228332 & 0.129952 \\ 
3.200000 & 0.063500 & 2.984240 & 0.281502 & 0.138266 \\ 
3.200000 & 0.064500 & 2.545250 & 0.284807 & 0.141217 \\ 
3.200000 & 0.065500 & 2.194860 & 0.194749 & 0.145616 \\ 
3.200000 & 0.066500 & 2.753030 & 0.259098 & 0.149937 \\ 
3.200000 & 0.067500 & 2.425180 & 0.280530 & 0.165001 \\ 
3.200000 & 0.068500 & 2.973450 & 0.381149 & 0.194633 \\ 
3.200000 & 0.069500 & 1.674630 & 0.171965 & 0.184622 \\ 
\hline
\end{tabular}
  \end{center}
\end{table}

\begin{table}[p]
  \scriptsize

  \begin{center}
  \begin{tabular}{ccccc}

  \hline
    \hline
    $z$ & $k\: {\rm [s\,km^{-1}]}$ &  $P_{1D}(z,k)\: {\rm [km\,s^{-1}]}$ &  $\sigma_{\rm stat}(z,k)\: {\rm [km\,s^{-1}]}$ & $P^{noise}(z,k)\: {\rm [km\,s^{-1}]}$ \\
\hline
 3.555000 & 0.001500 & 88.865100 & 6.515710 & 0.068276 \\ 
3.555000 & 0.002500 & 73.205800 & 5.585860 & 0.065727 \\ 
3.555000 & 0.003500 & 56.506800 & 4.059340 & 0.059281 \\ 
3.555000 & 0.004500 & 55.431500 & 4.706380 & 0.065773 \\ 
3.555000 & 0.005500 & 53.716600 & 3.723840 & 0.067768 \\ 
3.555000 & 0.006500 & 49.200200 & 4.171520 & 0.062244 \\ 
3.555000 & 0.007500 & 40.856000 & 2.989860 & 0.063620 \\ 
3.555000 & 0.008500 & 44.584000 & 3.431320 & 0.067218 \\ 
3.555000 & 0.009500 & 39.392600 & 2.903970 & 0.070289 \\ 
3.555000 & 0.010500 & 39.809200 & 3.131610 & 0.062548 \\ 
3.555000 & 0.011500 & 39.328800 & 3.592160 & 0.070704 \\ 
3.555000 & 0.012500 & 35.795000 & 2.874880 & 0.074327 \\ 
3.555000 & 0.013500 & 32.274600 & 2.456710 & 0.069139 \\ 
3.555000 & 0.014500 & 27.916700 & 1.928550 & 0.065362 \\ 
3.555000 & 0.015500 & 23.825100 & 1.920700 & 0.066460 \\ 
3.555000 & 0.016500 & 26.716800 & 2.137040 & 0.066389 \\ 
3.555000 & 0.017500 & 27.938800 & 2.204940 & 0.068883 \\ 
3.555000 & 0.018500 & 23.226400 & 1.576110 & 0.063949 \\ 
3.555000 & 0.019500 & 22.242000 & 1.652990 & 0.075736 \\ 
3.555000 & 0.020500 & 21.346200 & 1.681510 & 0.070140 \\ 
3.555000 & 0.021500 & 19.947400 & 1.620770 & 0.067363 \\ 
3.555000 & 0.022500 & 17.548600 & 1.215020 & 0.067391 \\ 
3.555000 & 0.023500 & 19.394300 & 1.439100 & 0.081993 \\ 
3.555000 & 0.024500 & 21.075800 & 1.770870 & 0.077001 \\ 
3.555000 & 0.025500 & 15.565100 & 1.461320 & 0.073399 \\ 
3.555000 & 0.026500 & 15.723200 & 1.105650 & 0.080061 \\ 
3.555000 & 0.027500 & 14.708600 & 1.155350 & 0.081257 \\ 
3.555000 & 0.028500 & 16.462900 & 1.432750 & 0.079846 \\ 
3.555000 & 0.029500 & 15.304700 & 1.261340 & 0.082697 \\ 
3.555000 & 0.030500 & 14.726000 & 1.016610 & 0.079283 \\ 
3.555000 & 0.031500 & 12.821200 & 1.061250 & 0.081678 \\ 
3.555000 & 0.032500 & 11.147600 & 0.855343 & 0.081194 \\ 
3.555000 & 0.033500 & 11.369800 & 0.947062 & 0.095574 \\ 
3.555000 & 0.034500 & 12.098200 & 0.959637 & 0.092569 \\ 
3.555000 & 0.035500 & 11.719500 & 0.915917 & 0.099837 \\ 
3.555000 & 0.036500 & 10.815400 & 0.776541 & 0.089857 \\ 
3.555000 & 0.037500 & 11.036900 & 0.844236 & 0.096920 \\ 
3.555000 & 0.038500 & 9.943440 & 0.820234 & 0.093338 \\ 
3.555000 & 0.039500 & 8.680970 & 0.709705 & 0.102320 \\ 
3.555000 & 0.040500 & 9.349400 & 0.737119 & 0.093777 \\ 
3.555000 & 0.041500 & 8.327400 & 0.612365 & 0.109957 \\ 
3.555000 & 0.042500 & 8.616410 & 0.729467 & 0.115910 \\ 
3.555000 & 0.043500 & 8.643760 & 0.759264 & 0.108632 \\ 
3.555000 & 0.044500 & 8.159150 & 0.595760 & 0.112995 \\ 
3.555000 & 0.045500 & 8.031460 & 0.575558 & 0.124681 \\ 
3.555000 & 0.046500 & 9.380490 & 0.772225 & 0.121156 \\ 
3.555000 & 0.047500 & 7.350790 & 0.568898 & 0.120322 \\ 
3.555000 & 0.048500 & 6.265940 & 0.496989 & 0.128234 \\ 
3.555000 & 0.049500 & 6.701070 & 0.475256 & 0.128688 \\ 
3.555000 & 0.050500 & 6.040380 & 0.457834 & 0.142447 \\ 
3.555000 & 0.051500 & 7.350400 & 0.607125 & 0.140783 \\ 
3.555000 & 0.052500 & 6.035630 & 0.508089 & 0.154362 \\ 
3.555000 & 0.053500 & 6.873980 & 0.558892 & 0.148096 \\ 
3.555000 & 0.054500 & 5.963720 & 0.451035 & 0.145561 \\ 
3.555000 & 0.055500 & 5.283380 & 0.442484 & 0.179433 \\ 
3.555000 & 0.056500 & 6.385500 & 0.672334 & 0.171652 \\ 
3.555000 & 0.057500 & 4.891520 & 0.500542 & 0.189406 \\ 
3.555000 & 0.058500 & 4.792700 & 0.408010 & 0.187159 \\ 
3.555000 & 0.059500 & 5.216490 & 0.583486 & 0.197799 \\ 
3.555000 & 0.060500 & 4.631330 & 0.509557 & 0.209080 \\ 
3.555000 & 0.061500 & 4.726460 & 0.451763 & 0.201643 \\ 
3.555000 & 0.062500 & 3.939570 & 0.449889 & 0.235744 \\ 
3.555000 & 0.063500 & 4.849110 & 0.508501 & 0.215283 \\ 
3.555000 & 0.064500 & 4.247100 & 0.536781 & 0.236235 \\ 
3.555000 & 0.065500 & 4.129100 & 0.409987 & 0.254957 \\ 
3.555000 & 0.066500 & 3.984800 & 0.551231 & 0.260904 \\ 
3.555000 & 0.067500 & 3.895950 & 0.474554 & 0.315798 \\ 
3.555000 & 0.068500 & 3.546240 & 0.398746 & 0.299962 \\ 
3.555000 & 0.069500 & 3.584180 & 0.400489 & 0.296902 \\
\hline
\end{tabular}
  \end{center}
\end{table}

 \begin{table*}[p]
 \scriptsize
  \begin{center}
  \begin{tabular}{ccccc}
  
  \hline
    \hline
    $z$ & $k\: {\rm [s\,km^{-1}]}$ &  $P_{1D}(z,k)\: {\rm [km\,s^{-1}]}$ &  $\sigma_{\rm stat}(z,k)\: {\rm [km\,s^{-1}]}$ & $P^{noise}(z,k)\: {\rm [km\,s^{-1}]}$ \\
\hline
3.925000 & 0.001500 & 133.465000 & 10.212600 & 0.084319 \\ 
3.925000 & 0.002500 & 100.990000 & 6.886760 & 0.090986 \\ 
3.925000 & 0.003500 & 91.277800 & 5.020310 & 0.079877 \\ 
3.925000 & 0.004500 & 84.812100 & 6.253830 & 0.094176 \\ 
3.925000 & 0.005500 & 67.483600 & 4.582350 & 0.086142 \\ 
3.925000 & 0.006500 & 69.705300 & 4.834860 & 0.088004 \\ 
3.925000 & 0.007500 & 66.546700 & 4.333430 & 0.084770 \\ 
3.925000 & 0.008500 & 55.040100 & 5.287900 & 0.084694 \\ 
3.925000 & 0.009500 & 52.586500 & 3.739190 & 0.084841 \\ 
3.925000 & 0.010500 & 46.361200 & 2.959380 & 0.081574 \\ 
3.925000 & 0.011500 & 42.759000 & 2.978470 & 0.093503 \\ 
3.925000 & 0.012500 & 44.310500 & 3.276610 & 0.085221 \\ 
3.925000 & 0.013500 & 40.173000 & 2.449020 & 0.097450 \\ 
3.925000 & 0.014500 & 41.588900 & 2.872700 & 0.082236 \\ 
3.925000 & 0.015500 & 41.695000 & 2.685520 & 0.088972 \\ 
3.925000 & 0.016500 & 38.910000 & 2.747650 & 0.084902 \\ 
3.925000 & 0.017500 & 31.174100 & 2.605840 & 0.094354 \\ 
3.925000 & 0.018500 & 31.451900 & 2.136030 & 0.089190 \\ 
3.925000 & 0.019500 & 31.084700 & 2.246470 & 0.090757 \\ 
3.925000 & 0.020500 & 28.402200 & 2.053140 & 0.096776 \\ 
3.925000 & 0.021500 & 33.085800 & 2.453990 & 0.086740 \\ 
3.925000 & 0.022500 & 24.056600 & 1.517200 & 0.099324 \\ 
3.925000 & 0.023500 & 26.755200 & 2.163250 & 0.092142 \\ 
3.925000 & 0.024500 & 29.758100 & 2.442720 & 0.103604 \\ 
3.925000 & 0.025500 & 23.258200 & 1.831560 & 0.092181 \\ 
3.925000 & 0.026500 & 21.466500 & 1.531250 & 0.094768 \\ 
3.925000 & 0.027500 & 25.425000 & 2.045770 & 0.101195 \\ 
3.925000 & 0.028500 & 20.605800 & 1.505640 & 0.096641 \\ 
3.925000 & 0.029500 & 22.769000 & 1.629400 & 0.095187 \\ 
3.925000 & 0.030500 & 17.080000 & 1.134630 & 0.102865 \\ 
3.925000 & 0.031500 & 19.450900 & 1.575840 & 0.105337 \\ 
3.925000 & 0.032500 & 17.937600 & 1.337740 & 0.094964 \\ 
3.925000 & 0.033500 & 16.596200 & 1.281680 & 0.111852 \\ 
3.925000 & 0.034500 & 14.830300 & 1.034130 & 0.103423 \\ 
3.925000 & 0.035500 & 15.013800 & 1.149150 & 0.101046 \\ 
3.925000 & 0.036500 & 12.979600 & 0.865509 & 0.099199 \\ 
3.925000 & 0.037500 & 15.232900 & 1.215290 & 0.110191 \\ 
3.925000 & 0.038500 & 13.161600 & 1.025750 & 0.117859 \\ 
3.925000 & 0.039500 & 15.046400 & 1.087180 & 0.109847 \\ 
3.925000 & 0.040500 & 12.725100 & 0.849439 & 0.110580 \\ 
3.925000 & 0.041500 & 13.703400 & 1.072110 & 0.111636 \\ 
3.925000 & 0.042500 & 11.954200 & 0.908273 & 0.114444 \\ 
3.925000 & 0.043500 & 12.511800 & 0.954848 & 0.108401 \\ 
3.925000 & 0.044500 & 11.528500 & 0.854209 & 0.121583 \\ 
3.925000 & 0.045500 & 11.792000 & 1.066600 & 0.115582 \\ 
3.925000 & 0.046500 & 10.402500 & 0.729569 & 0.111408 \\ 
3.925000 & 0.047500 & 9.969150 & 0.654077 & 0.119368 \\ 
3.925000 & 0.048500 & 10.472600 & 0.800422 & 0.125843 \\ 
3.925000 & 0.049500 & 11.162100 & 0.925858 & 0.135256 \\ 
3.925000 & 0.050500 & 8.240230 & 0.689975 & 0.113853 \\ 
3.925000 & 0.051500 & 8.748400 & 0.679270 & 0.136302 \\ 
3.925000 & 0.052500 & 8.837000 & 0.768439 & 0.136001 \\ 
3.925000 & 0.053500 & 7.201600 & 0.581492 & 0.138900 \\ 
3.925000 & 0.054500 & 8.704030 & 0.620238 & 0.133671 \\ 
3.925000 & 0.055500 & 7.904330 & 0.652731 & 0.136839 \\ 
3.925000 & 0.056500 & 6.549770 & 0.499559 & 0.147127 \\ 
3.925000 & 0.057500 & 7.084960 & 0.507684 & 0.139924 \\ 
3.925000 & 0.058500 & 8.200610 & 0.544680 & 0.141880 \\ 
3.925000 & 0.059500 & 7.005290 & 0.631734 & 0.152421 \\ 
3.925000 & 0.060500 & 5.886190 & 0.486973 & 0.162624 \\ 
3.925000 & 0.061500 & 6.457330 & 0.425875 & 0.142332 \\ 
3.925000 & 0.062500 & 5.766620 & 0.478320 & 0.166818 \\ 
3.925000 & 0.063500 & 6.002120 & 0.441410 & 0.162566 \\ 
3.925000 & 0.064500 & 6.352780 & 0.556433 & 0.160762 \\ 
3.925000 & 0.065500 & 4.605540 & 0.358773 & 0.169277 \\ 
3.925000 & 0.066500 & 5.910770 & 0.469332 & 0.172540 \\ 
3.925000 & 0.067500 & 5.466390 & 0.465330 & 0.188030 \\ 
3.925000 & 0.068500 & 4.602840 & 0.294872 & 0.167992 \\ 
3.925000 & 0.069500 & 4.246550 & 0.339262 & 0.194429 \\ 
 \hline
\end{tabular}
  \end{center}
\end{table*}

\end{document}